\documentclass[%
reprint,
superscriptaddress,
amsmath,amssymb,
aps,
prl,
floatfix,
]{revtex4-2}

\usepackage{graphicx} 
\usepackage{dcolumn} 
\usepackage{bm} 
\usepackage{bbm}
\usepackage{braket}
\usepackage{comment}
\usepackage[dvipsnames]{xcolor}

\usepackage{lipsum}

\usepackage{booktabs}

\definecolor{darkblue}{rgb}{0.,0.,0.6}
\definecolor{darkgreen}{rgb}{0.,0.6,0.0}
\definecolor{darkorange}{rgb}{0.8,0.4,0.2}

\newcommand{\Yb}{$^{171}$Yb$^+\,$}
\newcommand{\Ybc}{$^{172}$Yb$^+\,$}
\usepackage{hyperref} 

\begin{document}
\preprint{APS/123-QED}
\raggedbottom

\title{Experimental Realization of Thermal Reservoirs with Tunable Temperature in a Trapped-Ion Spin-Boson Simulator}

\author{Visal So}
\email{vs39@rice.edu}
\affiliation{Department of Physics and Astronomy and Smalley-Curl Institute, Rice University, Houston, TX 77005, USA}
\author{Mingjian Zhu}
\affiliation{Department of Physics and Astronomy and Smalley-Curl Institute, Rice University, Houston, TX 77005, USA}
\author{Midhuna Duraisamy Suganthi}
\affiliation{Department of Physics and Astronomy and Smalley-Curl Institute, Rice University, Houston, TX 77005, USA}
\affiliation{Applied Physics Graduate Program, Smalley-Curl Institute, Rice University, Houston, TX 77005, USA }
\author{Abhishek Menon}
\affiliation{Department of Physics and Astronomy and Smalley-Curl Institute, Rice University, Houston, TX 77005, USA}
\author{George Tomaras}
\affiliation{Department of Physics and Astronomy and Smalley-Curl Institute, Rice University, Houston, TX 77005, USA}
\affiliation{Applied Physics Graduate Program, Smalley-Curl Institute, Rice University, Houston, TX 77005, USA }
\author{Roman Zhuravel}
\affiliation{Department of Physics and Astronomy and Smalley-Curl Institute, Rice University, Houston, TX 77005, USA}
\author{Han Pu}
\affiliation{Department of Physics and Astronomy and Smalley-Curl Institute, Rice University, Houston, TX 77005, USA}
\author{Guido Pagano}
\email{pagano@rice.edu}
\affiliation{Department of Physics and Astronomy and Smalley-Curl Institute, Rice University, Houston, TX 77005, USA}

\begin{abstract}
    We propose and demonstrate an experimental scheme to engineer thermal baths with independently tunable temperatures and dissipation rates for the motional modes of a trapped-ion system. This approach enables robust thermal-state preparation and quantum simulations of open-system dynamics in bosonic and spin-boson models at well-controlled finite temperatures. We benchmark our protocol by experimentally realizing out-of-equilibrium dynamics of a charge-transfer model at different temperatures. We observe that, when the process occurs at a higher temperature, the transfer rate spectrum broadens, with reduced rates at small donor-acceptor energy gaps and enhanced rates at large gaps. We then employ our scheme to study local-temperature effects in a two-mode vibrationally assisted exciton transfer system, where we observe thermally activated interference pathways for excitation transfer.
\end{abstract}

\maketitle

\emph{Introduction} -- The competition between Coulomb interactions and trapping potentials in trapped-ion crystals gives rise to effective harmonic oscillator modes that describe the collective ion motion. Although these modes can, in principle, serve as native degrees of freedom for quantum applications, complementing the internal electronic states of the ions, they have more commonly been used as mediators for qubit-qubit interactions \cite{monroe2021programmable}. Consequently, experimental efforts have traditionally focused on keeping these bosonic degrees of freedom as close to the ground state as possible by optimizing cooling techniques and mitigating heating mechanisms to minimize temperature-related errors in quantum computing and simulation \cite{wineland1997issue,myatt2000decoherence,brownnutt2015noise}. However, the use of trapped-ion bosonic modes as information carriers and as active degrees of freedom in purely bosonic and spin-boson models has recently attracted growing interest \cite{chen2021bosonic,pagano2025varenna}. Non-classical motional states in trapped-ion systems have been used to encode continuous-variable logical states, such as Gottesman-Kitaev-Preskill (GKP) states \cite{fluhmann2019encoding,matsos2024gkp} and localized multi-component Schrödinger’s cat states \cite{rojkov2024stabilizationcatstatemanifoldsusing}, which are promising tools for efficient error correction \cite{gkp2001original,michael2016ecc,grimsmo2020bosoniccode}. Moreover, by engineering phonon-phonon couplings mediated by the ions’ spin degrees of freedom or employing the anharmonicity of the Coulomb potential, the harmonic oscillator quanta themselves can act as computational units, enabling boson sampling and simulations of purely bosonic Hamiltonians \cite{toyoda2015hom,debnath2018boson,zhang2018NOON,ding2018trilinear,katz2023boson,hou2024modulate}.

In addition, exploiting both spin and bosonic degrees of freedom on equal footing allows us to efficiently simulate spin-boson systems through spin-phonon digital \cite{zohreh2021spinphonon,than2025spinphonon} or analog protocols \cite{zahringer2010walk,clos2016thermalization,lv2018rabi,cai2021rabi,mei2022rabihubbard,muralidharan2023jch}. In the analog case, native interactions between the ions’ degrees of freedom and controlled electromagnetic fields are used to engineer the spin-boson Hamiltonians of interest, including linear vibronic coupling (LVC) models, which can be used to study key chemical dynamics, such as charge transfer and dissociation reactions \cite{gorman2018VAET,valahu2023conical,whitlow2023conical,macdonell2023timedomain}.
Combined with reservoir-engineering techniques, this approach extends naturally to non-equilibrium, open-system dynamics, where environmental effects are non-negligible \cite{kang2024chemical,olaya-agudelo2025open,pagano2025varenna}. Therefore, thermal-reservoir engineering with trapped ions, both for thermal-state preparation and for probing open-system dynamics, can serve as a valuable resource for quantum applications. For example, models of many chemical reactions assume that the system and its surroundings are in global thermal equilibrium \cite{laidler1987kinetics,truhlar1996TST,petruccione2007}. Consequently, the ability to establish a finite-temperature reservoir paves the way to experimentally study these processes under realistic thermodynamic conditions. Moreover, such reservoirs enable experimental studies of thermal entanglement and the temperature dependence of decoherence processes \cite{fedortchenko2014finitetemp,wu2017coherence}. They also provide a platform for investigating thermalization and heat flow across temperature gradients \cite{lin2011thermalization}. Existing thermal-reservoir engineering techniques typically rely on cooling processes \cite{poyatos1996cooling,so2024electrontransfer,so2025multimode,kienzler2015thesis,cetina2022cooling}, which pump the system into the vacuum state \cite{Monroe1995a} or into squeezed and entangled states \cite{carvalho2001stateprotect,kienzler2015reservoir,zhu2025dissipation,li2025entanglement}, or on noise injection \cite{myatt2000decoherence,navickas2025chemical} and stochastic processes \cite{sun2025quantum}, which yield infinite-temperature steady states. By contrast, controlled competition between heating and cooling processes has been employed to study phase transitions in the trapped ions’ motional state \cite{behrle2023phononlaser} and to realize quantum van der Pol oscillators using combinations of linear and non-linear dissipation channels \cite{li2025vanderpol}.

In this work, we propose a method that combines controlled heating and cooling of trapped ions’ motional modes to engineer thermal baths with independently tunable temperature and dissipation rate, an essential feature for faithfully modeling most chemical and physical systems. Experimentally, we induce motional excitation by broadcasting electric-field signal with stochastic phases and remove phonon energy via laser cooling. We test this scheme on a dual-species ion chain to probe temperature effects in a simplified charge-transfer model, observing a broadening of the transfer rate spectrum with respect to the donor-acceptor energy gap at higher temperatures. We also explore the role of local temperature in a two-mode system, where vibrationally assisted transfer exhibits thermal activation of coherent transfer pathways. Our work provides a key-enabling tool for studying open quantum systems under the influence of thermal environments, with direct relevance to real chemical systems (e.g., proton-coupled electron transfer in catalysis \cite{mayer2004proton}), biological systems (e.g., exciton transfer in photosynthetic complexes \cite{fassioli2014vibration}), and physical systems (e.g., phonon-induced decoherence in solid-state qubits \cite{jarmola2012diamond}) as well as for dissipative thermal-state preparation, useful for applications in quantum information science.

\emph{Theory} -- Throughout this work, we adopt the convention of setting $\hbar=1$. By simultaneously implementing cooling and heating processes on any chosen motional mode of a trapped-ion system at rates $\gamma_c$ and $\gamma_h$, respectively, the dynamics of the system is given by:
\begin{eqnarray}
    &&\frac{\partial\rho}{\partial t}\!=-i[H,\rho] +\gamma_c (n_{\rm ss}\!+\!1)\mathcal{D}_{a}[\rho] + \gamma_c n_{\rm ss} \mathcal{D}_{a^\dagger}[\rho],\;\;\;\;
    \nonumber\\
    &&\mathcal{D}_c[\rho]=c\rho c^\dagger -\frac{1}{2}\{c^\dagger c,\rho\},
    \label{eq_master}
\end{eqnarray}
where $H$ is the total Hamiltonian of the system, $n_{\rm ss}=\gamma_h/\gamma_c$, and $a$ and $a^\dagger$ represent the annihilation and creation operators, respectively. Considering only the dissipation terms, the steady state of the motional mode is a thermal state characterized by $n_{\rm ss}\equiv\bar{n}$ \cite{carmichael2013statistical} (see Supplemental Material \cite{so2025fT_supp} for a detailed proof). Therefore, the dissipation rate can be controlled by $\gamma_c$, while the bath temperature can be set independently by tuning $\gamma_h$.  
To benchmark the effect of the thermal bath, it is convenient to track the time evolution of the bosonic number operator $n=a^\dagger a$. Using Eq.~\eqref{eq_master}, the corresponding Heisenberg equation for $n$ can be derived when $[H,\rho]=0$, yielding the following solution:
\begin{equation}
    \braket{n(t)} = n_{\rm ss} + (\braket{n(t=0)}- n_{\rm ss})e^{-\gamma_c t}.
    \label{eq_ndynamics}
\end{equation}
Since Eq.~\eqref{eq_master} applies to any motional mode undergoing cooling and heating, independently controlling these processes on several modes allows us to simultaneously realize engineered baths for multiple motional modes, each with an independently tunable temperature.

\begin{figure}[!t]
\includegraphics[width=0.48\textwidth]{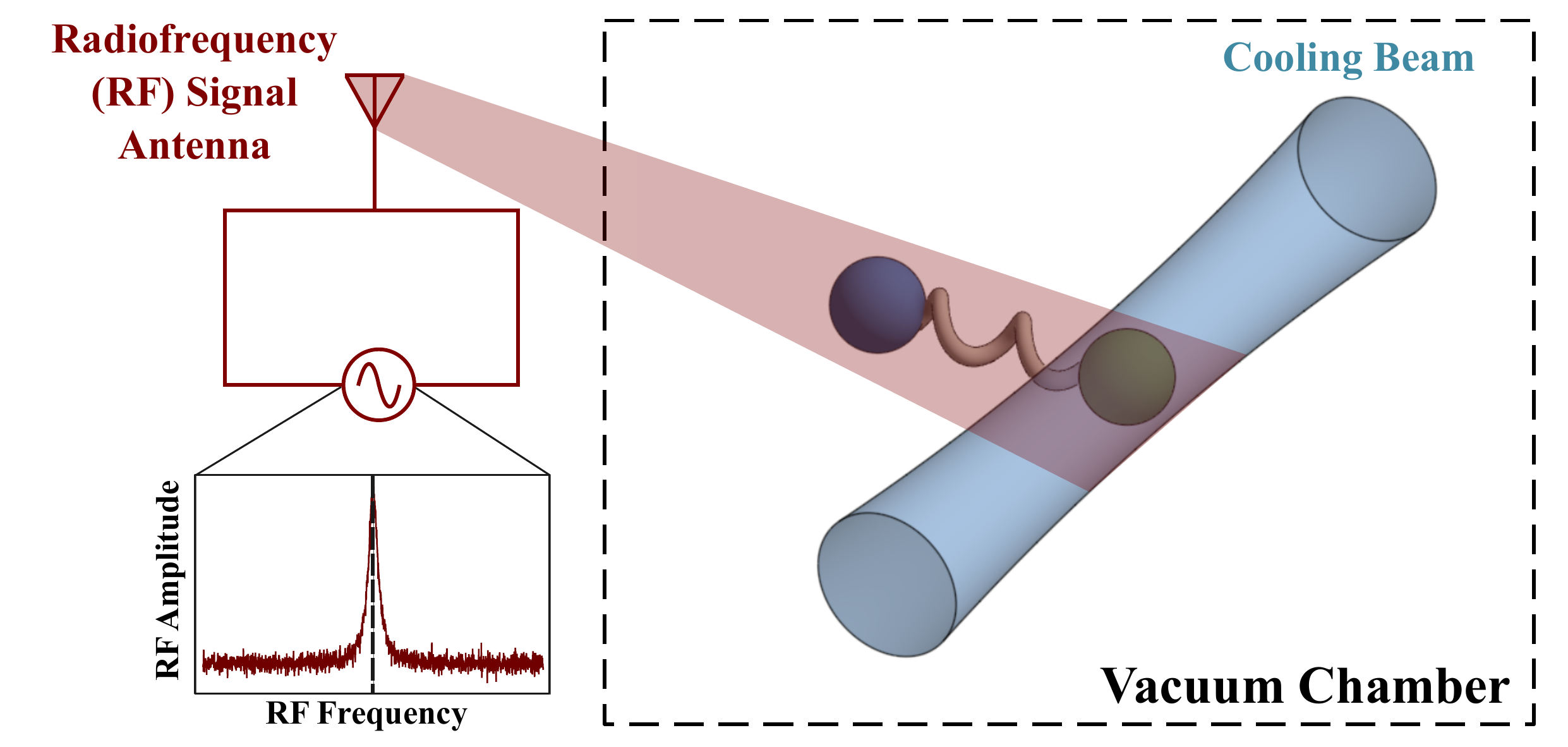}
\vspace{-1.5em}
\caption{{\bf Experimental setup for engineering a thermal reservoir with two trapped ions.} Independent control of the bath temperature and dissipation rate is achieved by simultaneously applying motional cooling and heating to the relevant mode. Coolant ion is addressed with a cooling beam that removes phonon excitations from the collective vibrational mode (connecting spring between the two spheres). It is not necessary to apply cooling beams to all ions, as long as the addressed ion participate in the targeted motional mode. Concurrently, an antenna broadcasts electric-field signal with stochastic phases at the motional-mode frequency from outside the vacuum chamber that hosts the ions to induce motional heating. This setup can be straightforwardly extended to realize multi-mode thermal baths in longer ion chains.}
\label{Fig_scheme}
\vspace{-1em}
\end{figure}
\emph{Experimental setup} -- A finite-temperature bosonic reservoir can be engineered by applying laser-cooling beams in combination with electric-field signals that induce motional heating of a specific mode of the ion chain (see Fig.~\ref{Fig_scheme} and End Matter). By applying multiple frequency tones that address different motional modes, this setup enables the engineering of multi-mode reservoirs with individually controllable temperatures. Experimentally, tunable cooling rates for selected motional modes are achieved by applying multiple frequencies in the laser-cooling beams, and injected electric-field signals with controlled-amplitude tones resonant with those modes are used for heating. The experimental configurations used in this work are described in the End Matter.

\begin{figure}[!t]
\includegraphics[width=0.48\textwidth]{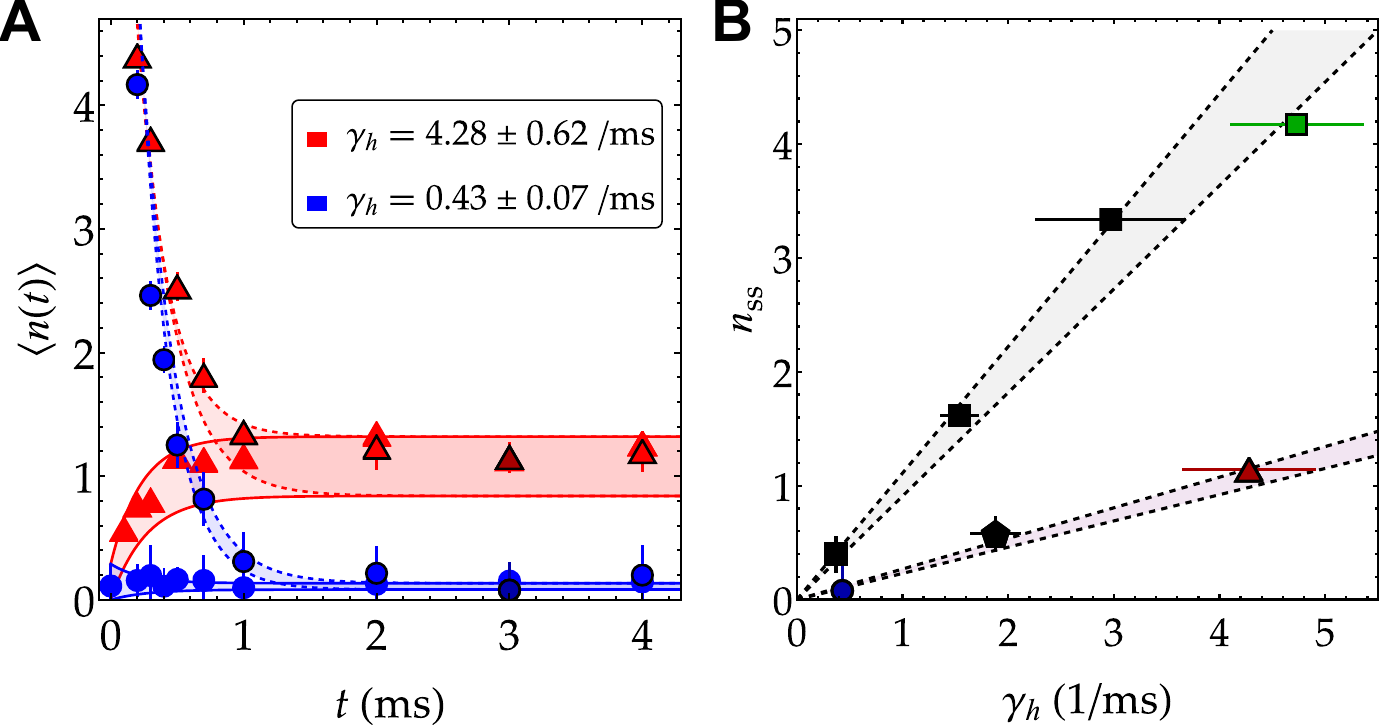}
\vspace{-2 em}
\caption{{\bf Thermal bath controls.} (A) Phonon-number dynamics from interacting with the engineered thermal bath. Red triangles and curves are associated with $\gamma_h=4.28\pm0.62$/ms, while blue circles and curves correspond to $\gamma_h=0.43\pm0.07$/ms. In both cases, $\gamma_c=4.03\pm 0.31$/ms. The curves are exponential functions with maximum and minimum time constants $1/\gamma_c$ and steady states given by $\braket{n(t\rightarrow\infty)}=\gamma_h/\gamma_c$. Data points with black outlines and dashed curves correspond to when the system starts from the Doppler-cooled temperature ($\braket{n(t=0)} \approx 7$), while data points with no outline and solid curves describe the realizations when the system is initialized from $\braket{n(t=0)}\approx0.1$. Darkened data points are also used in (B) (and Fig.~\ref{Fig_steadystate_20250818} of End Matter). (B) $n_{\rm ss}$ versus $\gamma_h$. Square data points correspond to the measured average phonon numbers after interacting with the engineered thermal bath, given by $\gamma_c = 1.00\pm0.10$/ms and $\gamma_h$, for 6 ms, while data points of other shapes are associated with the measured average phonon numbers after interacting with the engineered thermal bath, given by $\gamma_c=4.03\pm 0.31$/ms and $\gamma_h$, for 3 ms. The grey and purple bands are bounded by \{$\gamma_h/(\gamma_c=1.00-0.10$/ms), $\gamma_h/(\gamma_c=1.00+0.10$/ms)\} and \{$\gamma_h/(\gamma_c=4.03-0.31$/ms), $\gamma_h/(\gamma_c=4.03+0.31$/ms)\}, respectively. Colored data points are used in Fig.~\ref{Fig_steadystate_20250818} of End Matter.}
\label{Fig_dynamics_20250818}
\vspace{-1em}
\end{figure}

\emph{Controlling thermal reservoirs} -- In this section, we demonstrate the control of thermal baths with independently tunable temperature and dissipation rate by simultaneously applying motional heating and cooling to a center-of-mass (COM) mode of a single trapped \Yb ion. The average phonon-number evolution of this motional mode is expected to follow Eq.~\eqref{eq_ndynamics}, which we characterize experimentally as follows. We first prepare the motional mode to either $\braket{n(t=0)}\approx7$ via Doppler cooling or $\braket{n(t=0)}\approx0.1$ via resolved-sideband cooling, followed by a qubit-reset pulse to initialize the qubit in the $\ket{\downarrow}_z$ state. We then let the system evolve under the interaction with the engineered thermal bath described in Eq.~\eqref{eq_master} for time $t$. At the end of the evolution, we measure the phonon population $\braket{n(t)}$ by fitting the spin evolution under an anti-Jaynes-Cummings Hamiltonian with the first-order blue-sideband drive (see End Matter for details).
To measure the engineered $\gamma_h$ and $\gamma_c$, we independently capture the $\braket{n(t)}$ evolutions under heating and cooling, respectively, following the same analysis used for their combined action in our thermal-bath engineering scheme. We then perform a linear fit to extract $\gamma_h$ and an exponential fit to extract $\gamma_c$ (see Supplemental Material \cite{so2025fT_supp}). Knowing $\gamma_h$ and $\gamma_c$ allows us to predict $\bar{n}=n_{\rm ss}$ and control it accordingly.

In Fig.~\ref{Fig_dynamics_20250818}A, we show that the measured average phonon-number dynamics follow Eq.~\eqref{eq_ndynamics} for two engineered temperatures, starting from both a very low temperature ($\braket{n(t=0)}\approx 0.1$) and a Doppler-cooled temperature ($\braket{n(t=0)}\approx 7$). We also observe that the average phonon numbers at long times (steady states) fall within the estimated range of $n_{\rm ss} = \gamma_h/\gamma_c$ (indicated by the light red and blue bands for the higher- and lower-temperature cases, respectively). Moreover, we demonstrate control over the bath temperature, characterized by $n_{\rm ss}$, from near zero ($\approx 0.1$, limited by the intrinsic heating rate of the motional mode) up to about 4 by varying the motional heating rate while independently controlling two different cooling rates (see Fig.~\ref{Fig_dynamics_20250818}B). We emphasize that our proposed scheme is not limited to this temperature range. For example, higher heating rates or reduced cooling rates could be applied to achieve higher temperatures. Importantly, we also show that the phonon-population probabilities in the steady states, fitted as free parameters, closely follow those of thermal states (see End Matter).

\emph{Finite-temperature charge transfer} -- The ability to independently control the bath temperature and dissipation rate enables a wide range of applications in quantum information science, particularly robust state preparation and engineered reservoirs for open-system quantum simulation. To showcase both capabilities, we investigate the non-equilibrium transfer dynamics of an electron between a donor electronic site and an acceptor electronic site that are collectively coupled to a damped vibrational mode held at a specific temperature. This model was experimentally realized in Ref.~\cite{so2024electrontransfer} with tunable dissipation but without bath temperature control.
Here, the addition of controlled heating allows us to tune the bath temperature while retaining independent control over the dissipation rate.

A charge-transfer system can be minimally described by a LVC Hamiltonian \cite{schlawin2021electrontransfer,so2024electrontransfer}:
\begin{eqnarray}
H &=& \frac{\Delta E}{2}\sigma_z + V \sigma_x + \frac{g}{2}\sigma_z (a^\dagger + a) + \omega a^\dagger a,
\label{eq_H}
\end{eqnarray}
where $\ket{D}\equiv\ket{\uparrow}_z$ and $\ket{A}\equiv\ket{\downarrow}_z$ represent the donor and acceptor electronic sites, respectively. Here, $\sigma_{x,z}$ are the Pauli operators, with $\Delta E$ and $V$ denoting the donor-acceptor energy gap and coupling strength. These sites are linearly coupled to a harmonic oscillator with vibrational energy $\omega$, described by the creation ($a^\dagger$) and annihilation ($a$) operators, at a vibronic coupling strength $g$. Without electronic coupling, the vibronic system with $g\gtrsim\omega$ can be visualized as two displaced potential-energy surfaces with quantized energy levels: the donor surface is centered at $-g/2\omega$ and the acceptor surface at $+g/2\omega$ along the effective reaction coordinate, given by the position operator $y = y_0(a^\dagger+a)/2$, with $y_0 = \sqrt{1/2m\omega}$ and $m$ being the particle mass \cite{schlawin2021electrontransfer,so2024electrontransfer}.

When the electronic degree of freedom is strongly mixed ($V \sim g^2/4\omega$), the donor and acceptor energy surfaces hybridize into upper and lower adiabatic surfaces. This hybridization gives rise to eigenstates that are superpositions of donor and acceptor vibronic states, a hallmark feature of the strongly adiabatic regime of charge transfer. Since analytically describing the transfer dynamics in this regime is challenging due to the lack of a perturbative parameter in the system, it is of particular interest for experimental investigation. Naturally, charge transfer occurs in condensed-phase environments \cite{kang2024chemical}, which can be modeled as a thermal reservoir. Under the conditions $\gamma \ll \omega$ and $\gamma \ll k_B T$, this reservoir can be described by Eq.~\eqref{eq_master} (with $\gamma\equiv\gamma_c$ and $\bar{n}\equiv n_{\rm ss}$), where the bath temperature is given by $k_B T \approx \omega/\log(1+1/\bar{n})$ \cite{lemmer2018engineering,schlawin2021electrontransfer,so2024electrontransfer}. The vibrational spectrum relevant to condensed-phase charge transfer, such as in Cytochrome proteins \cite{sun2014lowfreq}, chlorophyll-based complexes \cite{ratsep2009chromo}, and bacteriochlorophyll-containing systems \cite{zhang2023pnas}, includes low-frequency modes with $\omega$ spanning tens to a few hundred cm$^{-1}$, and higher-frequency intramolecular modes extending up to 1000 cm$^{-1}$. At room temperature ($T\simeq 300$ K), vibrational modes in the hundred cm$^{-1}$ range correspond to thermal occupations $\bar{n}\simeq0.01-1$, which is neither negligible nor large, making finite-temperature effects significant. This range of vibrational frequencies is particularly relevant for energy transfer in molecular aggregates and light-harvesting materials \cite{fassioli2014vibration}, where the Born-Oppenheimer approximation breaks down because the vibrational and electronic energy scales are of the same order.

To jointly quantify the rate of dynamical equilibration and the amount of transferred population, we measure the inverse lifetime of the excitation residing in the donor site during the charge-transfer process \cite{skourtis1992photosynthesis,schlawin2021electrontransfer,so2024electrontransfer,so2025multimode}, $k_T=\frac{\int_{0}^{t_\text{sim}} P_D(t)dt}{\int_{0}^{t_\text{sim}} t P_D(t)dt} - \frac{2}{t_\text{sim}},$
where $t_{\rm sim}$ is the simulation duration, and $P_D(t)=(\braket{\sigma_z(t)}+1)/2$ \cite{so2024electrontransfer,so2025multimode,padilla2025delocalizedexcitationtransferopen} (see Supplemental Material \cite{so2025fT_supp}).

\begin{figure}[t!]
\includegraphics[width=0.45\textwidth]{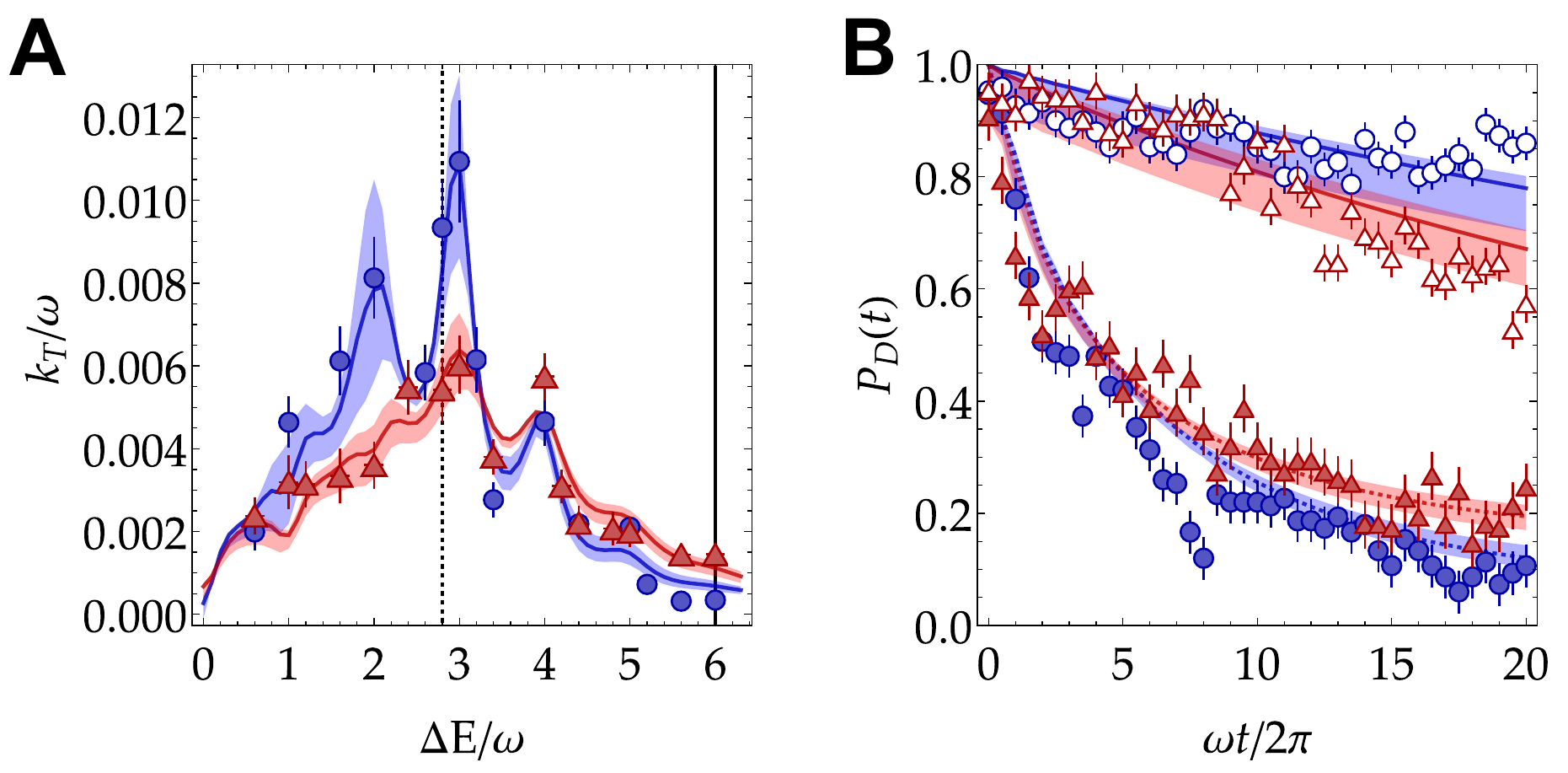}
\vspace{-1em}
\caption{{\bf Finite-temperature charge transfer.}
(A) Transfer rate versus donor-acceptor energy gap. Blue and red solid lines show the theoretical predictions of the transfer rate with $(V,\;g)=(0.2,\;1.1)\omega$ in contact with thermal reservoirs characterized by $\bar{n}=0.15$ and 0.80, respectively, at a dissipation rate of $\gamma = 0.036\omega$. These solid lines also consider experimental imperfections with $(\gamma_z,\;\gamma_m)  = (0.0014,\;0.0160)\omega$ (see Supplemental Material \cite{so2025fT_supp}). Blue-filled circles and red-filled triangles represent the corresponding experimental results with error bars obtained from resampling \cite{so2024electrontransfer}. (B) Donor population dynamics. Blue curves and circles show theoretical and experimental results, respectively, for the system interacting with the $\bar{n}=0.15$ reservoir, while red curves and triangles correspond to the $\bar{n}=0.80$ reservoir. Dashed curves and filled data points are for $\Delta E = 2.8\omega$, and solid curves with open data points are for $\Delta E = 6.0\omega$. The shaded bands on the theoretical curves correspond to the mean uncertainty ($\pm\,0.20$) in the most sensitive parameter in our setup, the bath temperature, and imperfect initial-state preparation.}
\label{Fig_ET}
\vspace{-1em}
\end{figure}

In Fig.~\ref{Fig_ET}A, the blue curve and data points show that the low-temperature ($\bar{n}=0.15$) transfer-rate spectrum in the strongly adiabatic regime ($V \sim g^2/4\omega$) exhibits two distinct behaviors with respect to the donor-acceptor energy gap (the experimental sequence is described in the Supplemental Material \cite{so2025fT_supp}). At small energy offsets ($\Delta E < 2\omega$), the monotonically increasing region arises from strong electronic coupling to off-resonant hybridized states, while remaining limited by the dissipation rate. In this case, the transfer rates are approximately proportional to the dissipation rate \cite{schlawin2021electrontransfer,so2024electrontransfer}. When $\Delta E > 2\omega$, we experimentally observe resonant peaks at $\ell\omega$ for $\ell=3$ and 4. Here, the initially localized donor state overlaps strongly with eigenstates of the upper-hybridized surface, which results in population trapping \cite{schlawin2021electrontransfer,so2024electrontransfer,so2025multimode}. At these resonances, the trapped population is released during the dynamics, transferring from the upper- to the lower-hybridized surface. Simultaneously, dissipation in the presence of a low-temperature reservoir relaxes the wavepacket into lower-energy levels of the lower-hybridized surface that strongly overlap with the acceptor site and are no longer resonant with the eigenenergy levels of the upper-hybridized surface, eventually driving the system to the steady state (see End Matter). As $\Delta E$ increases, the transfer rates diminish due to the weak couplings between higher-energy hybridized states on the two surfaces, which are governed by the overlaps between their motional wavefunctions (namely, Franck-Condon factors), suppressing the transfer-rate resonances at $\Delta E \gtrsim 4\omega$.

When the bath temperature is increased to $\bar{n}=0.80$ (red curve and data points in Fig.~\ref{Fig_ET}A), the transfer rates decrease at small donor-acceptor energy gaps but become slightly enhanced at larger gaps. This overall broadening of the spectrum can be attributed to the redistribution of thermal population across the hybridized energy surfaces: at higher temperatures, more population accumulates in higher-energy states. Consequently, for charge-transfer processes with small donor-acceptor gaps, higher temperatures place the initial donor population in higher-energy, hybridized states on both adiabatic surfaces, leaving a large fraction on the donor site (see End Matter) and thereby reducing transfer rates. This observation provides a new insight into the slope of the monotonically increasing region of the transfer-rate spectrum, which appears to be temperature-dependent in addition to its proportionality with the dissipation rate, found in Refs.~\cite{schlawin2021electrontransfer,so2024electrontransfer}. On the other hand, when $\Delta E$ is sufficiently large ($\Delta E\gtrsim4\omega$), thermal excitations redistribute the initially trapped donor population across eigenstates of the upper adiabatic surface, which are coherently transferred to the lower adiabatic surface at the $\Delta E=\ell\omega$ resonances, where $\ell$ is an integer (see End Matter). For $g\gtrsim\omega$, the coupling rates for this process, determined by the Franck-Condon factors, increase at moderately high temperatures (see Supplemental Material \cite{so2025fT_supp}), leading to enhanced transfer rates. These effects have been predicted in more complex systems with long-range interactions \cite{padilla2025delocalizedexcitationtransferopen}. We note that the details of the temperature effects depend on the system parameters ($V$, $g$, and $\omega$). 

\begin{figure}[t!]
\includegraphics[width=0.45\textwidth]{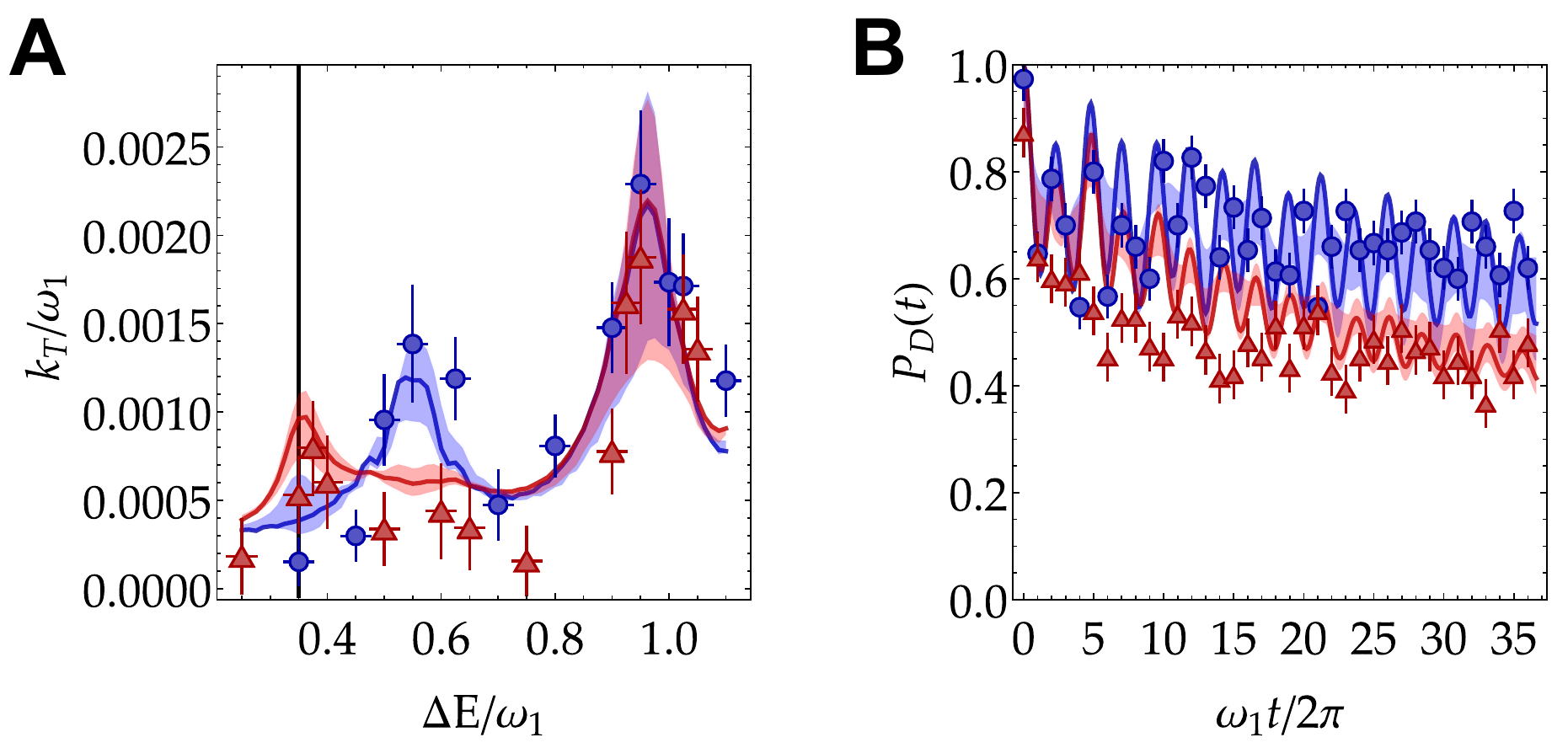}
\vspace{-1.2em}
\caption{{\bf Two-mode, vibrationally assisted exciton transfer at different local temperatures.} (A) Transfer rate versus donor-acceptor energy gap. Blue and red solid lines show the theoretical predictions of the transfer rate with $(V,\;\omega_2,\;g_1,\;g_2)=(0.14,\;0.60,\;0.33,\;0.18)\omega_1$ in contact with local thermal reservoirs characterized by $(\bar{n}_1,\bar{n}_2)=(0.10,\;0.02)$ and $(0.10,\;0.80)$, respectively, at dissipation rates $\gamma_1 = 0.013\omega_1$ and $\gamma_2 = 0.008\omega_2$.
These solid lines also consider experimental imperfections with $(\gamma_z,\;\gamma_m)  = (0.0004,\;0.0040)\omega_1$ (see Supplemental Material \cite{so2025fT_supp}). Blue-filled circles and red-filled triangles represent the corresponding experimental results with error bars obtained from resampling \cite{so2024electrontransfer}. (B) Donor population dynamics for $\Delta E = 0.35\omega_1$. Blue curves and circles show the theoretical and experimental results, respectively, for the system interacting with the $\bar{n}_2=0.02$ bath, while red curves and triangles correspond to the $\bar{n}_2=0.80$ bath. The shaded bands on the theoretical curves correspond to the mean uncertainty ($\pm\,0.20$) in the most sensitive parameters in our setup, the bath temperatures, and imperfect initial-state preparation.}
\label{Fig_thermallyact}
\vspace{-1em}
\end{figure}

\emph{Thermally activated transfer pathways in two-mode systems} -- We then simulate a LVC system featuring two dissipative vibrational modes ($\omega_1>
\omega_2$) in contact with thermal baths at different local temperatures and dissipation rates, described by \cite{so2025multimode}:
\begin{equation}
    H \!= V\sigma_x\! + \frac{\Delta E}{2}\sigma_z \!+ \sum_{i=1}^2\left\{\frac{g_i}{2}\sigma_z\!\left(a_i+a^\dag_i\right)\!+\omega_i a^\dag_i a_i\right\},
    \label{eq_H2M}
\end{equation}
\begin{equation}
    \frac{\partial\rho}{\partial t}\!=-i[H,\rho] +\sum_{i=1}^2\left\{\gamma_i (\bar{n}_i\!+\!1)\mathcal{L}_{a_i}[\rho] + \gamma_i \bar{n}_i \mathcal{L}_{a^\dagger_i}[\rho]\right\}.
    \label{eq_master2M}
\end{equation}
In the vibrationally assisted exciton transfer regime ($g_i\ll\omega_i$), vibrational modes primarily act as mediators of exothermic transfer, enabling energy exchange at discrete resonances $\Delta E \approx \sqrt{(\ell_1\omega_1 + \ell_2\omega_2)^2-(2V)^2}$, where $\ell_1$ and $\ell_2$ are integers \cite{so2025multimode}. For $0<\Delta E<\omega_1$, only two resonances appear in the transfer-rate spectrum at very low temperature (blue curve and data in Fig.~\ref{Fig_thermallyact}), near 0.53$\omega_1$ and 0.96$\omega_1$, each corresponding to single-phonon exchange between the electronic system and one of the two vibrational modes. As shown by the red curve and data, raising the local temperature of the $\omega_2$ mode selectively reduces the transfer rates associated with that mode’s energy exchange, since stronger temperature-induced couplings adversely keep more population on the donor site during the dynamics (see Supplemental Material \cite{so2025fT_supp}). Interestingly, a mixed-mode resonance near 0.29$\omega_1$ emerges, corresponding to excitation of the $\omega_1$ mode and de-excitation of the $\omega_2$ mode during coherent dynamics prior to equilibration. Despite the presence of dissipation, this transfer proceeds via interfering coherent pathways from combinations of phonon excitation and de-excitation. The blue curve and data show that this process is suppressed when both modes are globally very cold (see Supplemental Material \cite{so2025fT_supp} for explanation).

\emph{Outlook} -- In this work, we introduce a versatile and scalable method for engineering finite-temperature reservoirs for trapped-ion motion by balancing controlled heating induced by tailored broadcast electric-field signal with continuous resolved-sideband laser cooling. We experimentally demonstrate independent control over both bath temperatures and equilibration rates. Since high-temperature fits become increasingly sensitive to outlier populations and to drifts in motional mode frequencies or laser powers, we reconstruct thermal phonon distributions only up to $\bar{n}\lesssim 4$. We emphasize that this is an experimental, rather than fundamental, limitation, and that the accessible temperature range could be extended through improved technical stability, refined analysis, and improved heating and cooling protocols.

Applying this method to a dual-species trapped-ion setup, we realize charge-transfer dynamics across distinct thermal environments, gaining clearer insight into the role of temperature in the charge-transfer process in the strongly adiabatic regime. We observe that the temperature-induced distribution of excitation population across the mixed potential-energy landscape strongly shapes how the transfer-rate spectrum varies with the donor-acceptor energy gap. We also investigate local-temperature effects in two-mode, vibrationally assisted exciton transfer and find that local temperature can activate otherwise suppressed transfer pathways via constructive interference. These results confirm the role of temperature in LVC models and showcase our ability to tune the reservoir temperature in these open systems. This constitutes an important extension of previous trapped-ion LVC studies, which were restricted to either near-zero-temperature \cite{so2024electrontransfer,so2025multimode} or effectively infinite-temperature conditions \cite{sun2025quantum,navickas2025chemical}.

Our method provides a useful tool for robust thermal-state preparation and for exploring finite-temperature open-system dynamics in purely bosonic and spin-boson models. With these capabilities, we move one step closer to emulating realistic chemical reactions \cite{marcus1993electron,olaya-agudelo2025open}, probing thermal entanglement \cite{valido2013gaussian,chanda2018entanglement}, realizing high-efficiency quantum thermal machines \cite{rossnagel2016engine,maslennikov2019fridge}, and examining quantum thermoelectric processes \cite{roy2010thermoelectric,tejero2025asym} to better understand the underlying mechanisms in their natural settings, as well as in the crossover conditions between the quantum and classical regimes \cite{barbara1996et}. Furthermore, resonantly driving higher-order motional processes with electric-field and laser-cooling tones offers a simple extension of our method, opening avenues to non-linear reservoirs and to structured, non-Gaussian bosonic equilibrium states. With this extension, one can explore dissipative quantum phase transitions and non-equilibrium dynamics with non-linear decays \cite{shah2024dissipative,mylnikov2025twophonon,liu2025observationsynchronizationquantumvan}, as well as possibly stabilize non-classical states that enhance protocols for bosonic computation and simulation \cite{rojkov2024stabilizationcatstatemanifoldsusing,matsos2024gkp}. Lastly, our technique not only expands the toolbox for trapped-ion quantum technologies but may also find applications in other bosonic platforms, particularly superconducting circuits.

\emph{Acknowledgments} -- 
G.P. acknowledges support from the Welch Foundation Award (grant no. C-2154), the Office of Naval Research Young Investigator Program (grant no. N00014-22-1-2282), the NSF CAREER Award (grant no. PHY-2144910), and the Office of Naval Research (grant no. N00014-23-1-2665 and N00014-24-1-2593). We acknowledge that this material is based on work supported by the U.S Department of Energy, Office of Science, Office of Nuclear Physics under the Early Career Award (grant no. DE-SC0023806). The isotopes used in this research were supplied by the US Department of Energy Isotope Program, managed by the Office of Isotope R$\&$D and Production. H.P. acknowledges support from the NSF (grant no. PHY-2513089) and the Welch Foundation (grant no. C-1669).

\bibliographystyle{sty}
\bibliography{ref}

\section*{End Matter}
\label{sec:end_matter}

\subsection*{Methods}

In this work, we realize our proposed scheme by simultaneously performing continuous resolved-sideband cooling and inducing heating on a chosen motional mode. For the demonstration of thermal-bath control, we use the red-sideband drive via a 355 nm Raman transition together with a 370 nm optical pumping beam for cooling the motional mode of a single \Yb ion at the trap frequency of $\omega_c=2\pi\;\times$ 3.904 MHz (a center-of-mass (COM) radial mode) with an intrinsic heating rate of $0.43\pm0.07$/ms. The qubit used in this setup consists of the two hyperfine clock states of the \Yb electronic ground-state manifold: $\ket{^2S_{1/2}, F=1, m_F=0}\equiv\ket{\uparrow}_z$ and $\ket{^2S_{1/2}, F=0, m_F=0}\equiv\ket{\downarrow}_z$.
For the application of this technique in quantum simulation of open systems, we drive the red-sideband process using quadrupole $\ket{^2S_{1/2}}\rightarrow\ket{^2D_{3/2}}$ transitions at 435.5 nm, along with a 935 nm repumper beam, on an ancilla \Ybc ion cotrapped with the qubit \Yb ion. This results in sympathetic cooling of the shared motional mode at $\omega_t =2\pi\;\times$ 3.776 MHz (an out-of-phase radial mode) with an intrinsic heating rate of $0.03\pm0.01$/ms, corresponding to the axial trap frequency of $\omega_a\approx2\pi\;\times$ 0.99 MHz. The qubit ion is simultaneously addressed by 355 nm Raman beams to engineer the LVC Hamiltonians \cite{so2024electrontransfer,so2025multimode} (see Supplemental Material \cite{so2025fT_supp}). Together with motional heating, this enables us to investigate the effects of temperature on the out-of-equilibrium dynamics of excitation transfer models.

Motional heating is induced by driving the ion chain with a radiofrequency (RF) electric field at the selected motional-mode frequency. The RF signal is delivered via an antenna mounted near the vacuum system (Fig.~\ref{Fig_scheme}, see Supplemental Material for details \cite{so2025fT_supp}), and its phase is varied stochastically. The drive therefore produces coherent displacements of the addressed motional mode whose direction in motional phase space varies randomly in time, resulting in a random-walk trajectory. When averaged over many uncorrelated trajectories, this random walk yields phase-space diffusion that can be approximated as Markovian heating within a Lindblad master-equation framework \cite{sun2025quantum}. In practice, we implement this by applying the drive in many small time steps with independently randomized phases within each realization, and then averaging the resulting dynamics over many realizations. For the Lindblad approximation to be valid, the time step must be short compared to all other relevant dynamical timescales in the system \cite{sun2025quantum}. In Fig.~\ref{Fig_dynamics_20250818}, each data point is given by $40$ random phase sequences, where each sequence consists of multiple phase-varying steps with a fixed time step of 0.05 ms and is repeated 30 times. We note that for large $n_{\rm ss}$, more sequences are required to obtain a well-approximated thermal distribution of the phonon population, as discussed in Ref.~\cite{sun2025quantum}. For $n_{\rm ss}\approx4$, we use 60 sequences.

\begin{figure}[!t]
\includegraphics[width=0.35\textwidth]{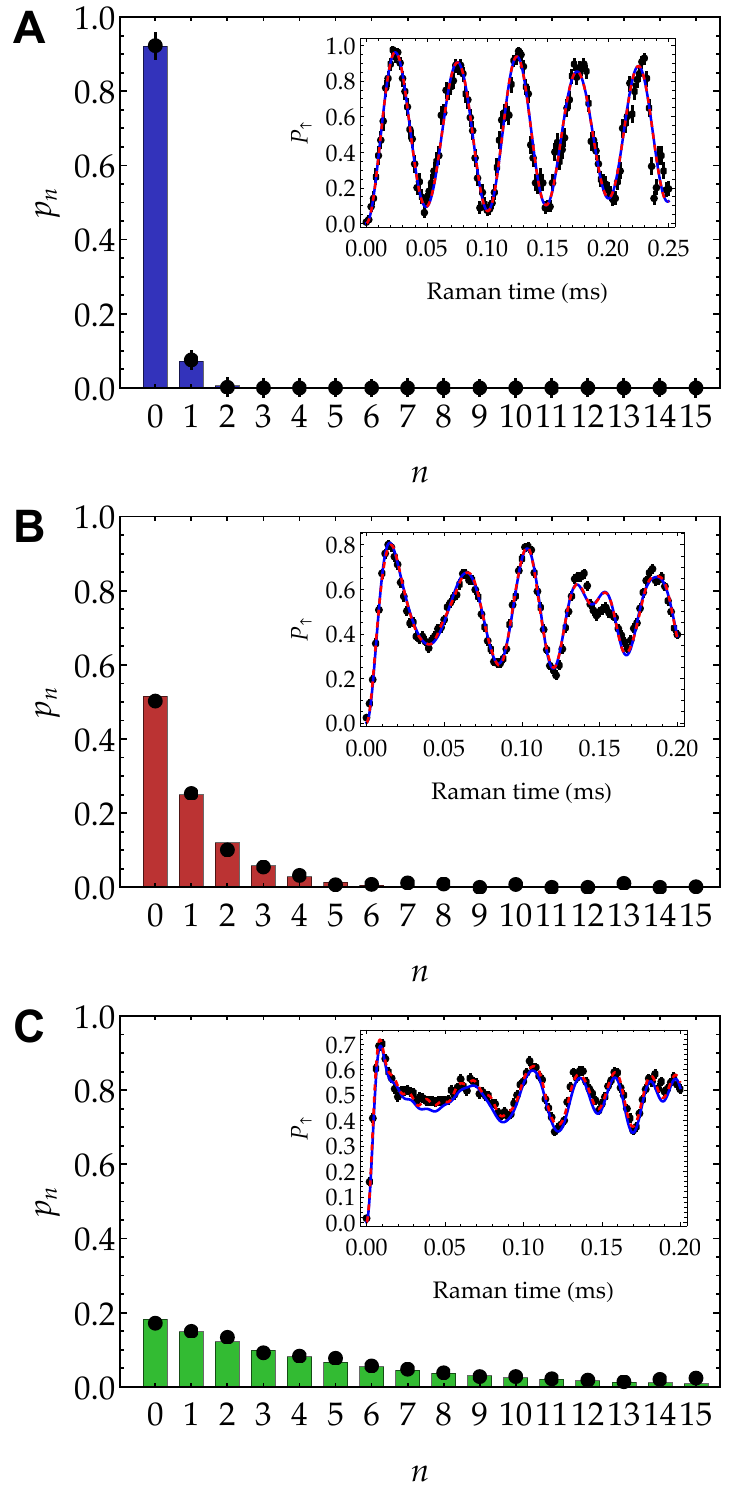}
\vspace{-1 em}
\caption{\textbf{Steady-state phonon-number distribution from the engineered thermal reservoir.} (A) $\gamma_c = 4.03\pm0.31$/ms and $\gamma_h = 0.43\pm0.07$/ms for 3 ms (dark blue in Fig.~\ref{Fig_dynamics_20250818} of the main text), (B) $\gamma_c = 4.03\pm0.31$/ms and $\gamma_h=4.28\pm0.62$/ms for 3 ms (dark red in Fig.~\ref{Fig_dynamics_20250818} of the main text), and (C) $\gamma_c = 1.00\pm0.10$/ms and $\gamma_h=4.73\pm0.62$/ms for 6 ms (green in Fig.~\ref{Fig_dynamics_20250818}B of the main text). Each inset corresponds to the spin dynamics of the steady-state system undergoing the probe blue-sideband drive for the phonon-number distribution measurement (black data points in the main figure). Bar charts represent the best-estimated thermal state bounded by $\gamma_h/\gamma_c$: (A) $0.09\pm0.02$, (B) $0.94\pm0.05$, and (C) $4.50\pm0.19$. Dashed red and solid blue curves in the insets describe the dynamics of the fitted phonon-number distribution and the estimated thermal state, respectively.}
\label{Fig_steadystate_20250818}
\end{figure}

We also find that, in the presence of motional dephasing, a stochastic-phase drive applied without temporal discretization can produce dynamics and steady states that are experimentally comparable to those predicted by the Lindblad thermal-bath master equation for relatively low temperature baths ($\bar{n}\lesssim 1$) when combined with cooling and averaged over many realizations (see Supplemental Material \cite{so2025fT_supp}). This approach enables higher data-taking rates and reduces calibration complexity, and we use it in the LVC-model simulations (Figs.~3 and 4).

Moreover, we note that motional heating can also be achieved either by directly injecting broadband noise into an ion-trap electrode \cite{myatt2000decoherence,navickas2025chemical} or by combining red- and blue-sideband optical drives with stochastic phases \cite{sun2025quantum,olaya-agudelo2025open}. A theoretical exploration of combining the latter method with qubit reset operations as an alternative approach for finite-temperature bath engineering is provided in the Supplemental Material \cite{so2025fT_supp}. Realizing a Markovian bath with broadband noise requires a correlation time much shorter than the relevant system timescales (i.e., a bandwidth much larger than the system’s characteristic frequencies), which can reduce mode selectivity by simultaneously driving multiple modes in mode-crowded chains. Therefore, this approach is better suited to global-heating protocols, where mode selectivity is not required. On the other hand, a stochastically varying-phase drive can preferentially heat a chosen mode, provided that the displacement drive’s Rabi coupling strength and detuning are smaller than the mode spacing. This makes the latter approach a scalable route to mode-selective heating.

To extract the phonon-number distribution of the motional state, we apply an anti-Jaynes-Cummings Hamiltonian using the first-order blue-sideband drive and fit the measured spin dynamics to:
\begin{equation}
    P_{\uparrow} (t)=\frac{1}{2}\sum_n p_n(t)\left[1-e^{-\gamma_{\rm d} t_{\rm p}}\cos{\left(\Omega t_{\rm p}\sqrt{n+1}\right)}\right],
    \label{eq_phonmeasure}
\end{equation}
where $p_n(t)$ is the population in Fock state $\ket{n}$ at the evolution time $t$, $\gamma_{\rm d}$ is an empirical decay constant given by spin and motional decoherences in the system, $t_{\rm p}$ is the duration of the blue-sideband drive, and $\Omega$ corresponds to the sideband Rabi frequency. From the fitted $p_n(t)$, we can calculate the average phonon excitation of the motional mode $\braket{n(t)}$, whose standard deviation is obtained from the covariance matrix of the fitting results \cite{cai2021rabi}. For the fits, we apply loose constraints to suppress extreme outliers that could distort the evaluation of $\braket{n(t)}$ (see Supplemental Material \cite{so2025fT_supp}).

\subsection*{Steady-state phonon-number distributions}

We show in Fig.~\ref{Fig_steadystate_20250818} that the phonon-population distributions of the steady states in Fig.~\ref{Fig_dynamics_20250818} of the main text resemble those of thermal states. We compare the free-fit estimates (black points) with the thermal distributions (colored bars), whose best-fit average phonon numbers are constrained by the measured values of $\gamma_h/\gamma_c$ and their uncertainties.

\subsection*{Adiabatic surfaces in charge transfer}
\begin{figure}[t!]
\includegraphics[width=0.45\textwidth]{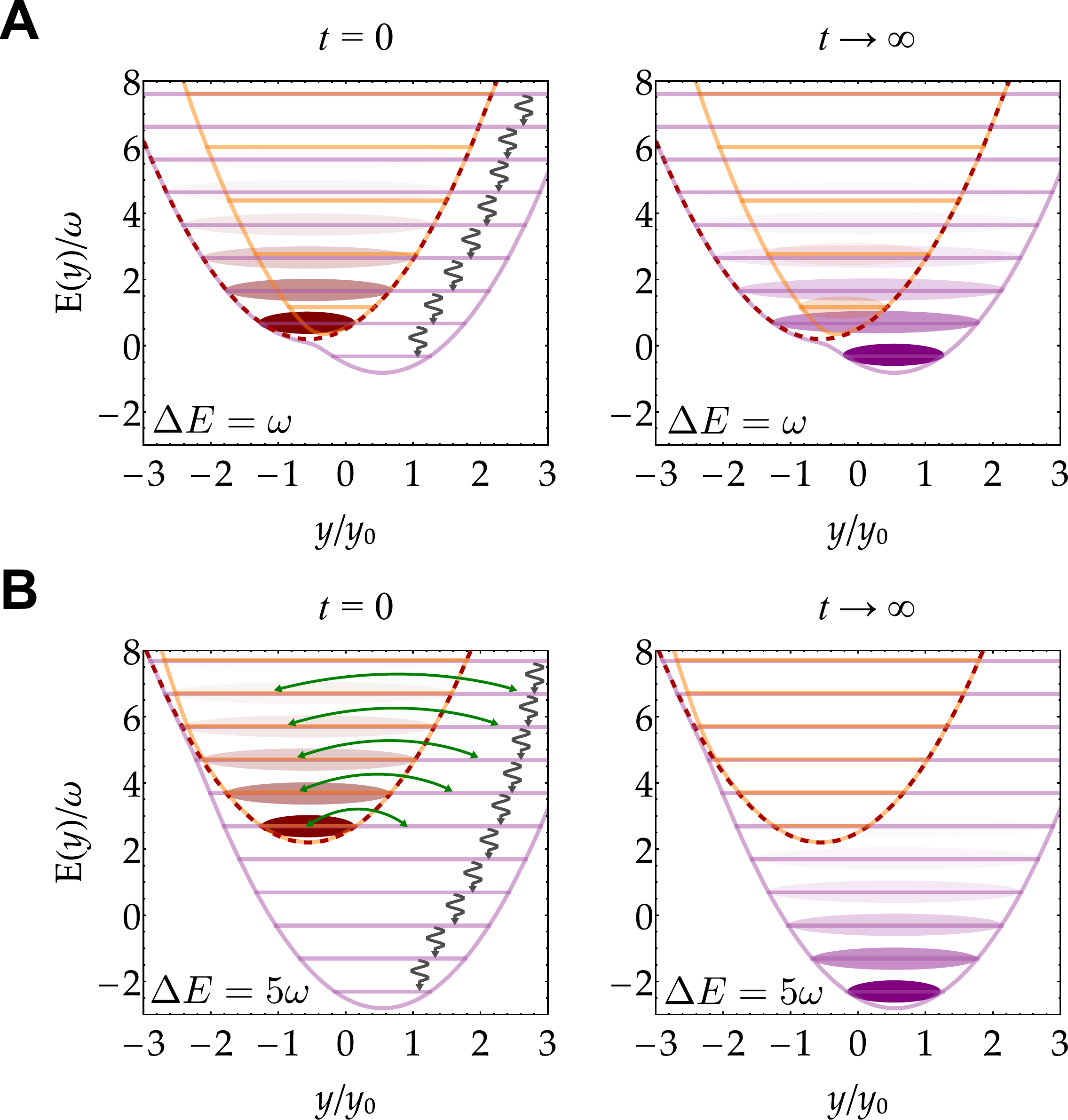}
\caption{{\bf Adiabatic surfaces in charge transfer.} (A) and (B) show the energy landscapes of the systems in Fig.~\ref{Fig_ET} of the main text, for a low energy gap $\Delta E =\omega$ and a high energy gap $\Delta E=5\omega$ at $\bar{n}=0.8$, respectively. Left subpanels correspond to the initial state and right to the steady state. Purple and orange parabolas (with horizontal lines) represent the lower and upper adiabatic surfaces (with their eigenenergies). Dashed red parabola is the uncoupled donor energy surface ($V=0$). For low $\Delta E$, charge transfer is primarily facilitated by coherent oscillations between the donor and acceptor sites within the adiabatic surfaces, while resonant transitions in the high $\Delta E$ case (green arrows) release the donor population initially trapped on the upper adiabatic surface into the lower adiabatic surface with a strong overlap with the acceptor site. During the dynamics, dissipation (grey arrows) suppresses coherence and leads to equilibration.}
\label{Fig_adiabatic}
\end{figure}

Visualizing the hybridized energy surfaces is helpful in understanding charge-transfer behaviors in the strongly adiabatic regime. Figs.~\ref{Fig_adiabatic}A-B show the adiabatic surfaces for the charge-transfer system in the main text at $\Delta E=\omega$ and $\Delta E = 5\omega$ with $\bar{n}=0.8$, respectively. The former represents the low-$\Delta E$ conditions, where transfer rates increase monotonically, while the latter clarifies the transfer resonances arising from initial population trapping.

As seen in Fig.~\ref{Fig_adiabatic}A, small $\Delta E$ leads to strong mixing of donor and acceptor sites on both adiabatic surfaces, especially at higher-energy hybridized levels. Thus, at elevated temperatures, the steady-state population occupies these higher-energy levels, leaving a large fraction on the donor site and thereby reducing the transfer rate. Conversely, for very large $\Delta E$, donor-acceptor mixing is minimal, and the upper and lower adiabatic surfaces closely approximate the uncoupled donor and acceptor surfaces (see Fig.~\ref{Fig_adiabatic}). In this regime, the initial donor population remains trapped unless resonant transfer from the upper to the lower adiabatic surface is activated (green arrows), with rates set by the inter-surface Franck-Condon factors. Although this process is coherent, dissipation from the bath stabilizes the excitation in the lower adiabatic levels at low and moderately high temperatures ($k_B T \sim \omega$). We discuss the very high-temperature (classical) regime with $k_B T\gg\omega$ in the Supplemental Material \cite{so2025fT_supp}.

\setcounter{figure}{0}
\renewcommand{\figurename}{Figure}
\renewcommand{\thefigure}{S\arabic{figure}}
\setcounter{equation}{0}
\renewcommand{\theequation}{S.\arabic{equation}}
\section*{Supplemental Material}

\subsection*{Steady-state phonon distribution}
\label{app_steady}

Regarding the steady-state phonon distribution of the bosonic degree of freedom of the trapped-ion system, described by Eq.~(1) in the main text \cite{so2025fT} with $[H,\rho]=0$, we can consider the diagonal component of the density matrix $p_n=\braket{n|\rho|n}$, which yields:
\begin{eqnarray}
    \frac{\partial p_n}{\partial t}&=&(\gamma_c+\gamma_h)(p_{n+1}(n+1)-p_n n) \nonumber \\
    &\quad&+\;\gamma_h(p_{n-1} n - p_{n}(n+1)),
\end{eqnarray}
where $\frac{\partial p_n}{\partial t}=0$. This equation gives the detailed-balance recurrence for $n\geq1$:
\begin{equation}
    (\gamma_c+\gamma_h) p_n = \gamma_h p_{n-1},
\end{equation}
resulting in:
\begin{equation}
    \frac{p_{n}}{p_{n-1}}=\frac{\gamma_h/\gamma_c}{\gamma_h/\gamma_c+1}=\frac{n_{\rm ss}}{n_{\rm ss}+1}.
\end{equation}
This condition satisfies the form of a thermal state characterized by the average occupation number $n_{\rm ss}$.

\vspace{0em}
\subsection*{Fitting constraints for phonon-number distribution} \label{app_constraint}
\vspace{0em}
Extreme outliers in $p_n$ from free-parameter fits to Eq.~(6) in the End Matter \cite{so2025fT} can affect the accuracy of the estimate of $\braket{n(t)}$. To mitigate this, we first estimate a thermal mean occupation number $n_{\rm ave}$ by fitting the dynamics to:
\begin{equation}
    P_{\uparrow} (t,t_{\rm p})=\frac{1}{2}\sum_n p_n^{\rm th}(t)[1-e^{-\gamma t_{\rm p}}\cos{( \Omega t_{\rm p}\sqrt{n+1})}],
    \label{eq_thermal}
\end{equation}
where $p_n^{\rm th}\equiv\frac{n_{\rm ave}^n}{(n_{\rm ave}+1)^{n+1}}$. We then use $p_n^{\rm th}$ to weakly constrain the free-parameter fits by enforcing $p_n<c\times p_n^{\rm th}$, where the constraint factor $c$ is chosen between 50 and 500. These constraints are relevant only at very low temperatures ($n_{\rm ss}<1$); at higher temperatures, the fitting parameters are essentially unconstrained. We also compare the fitted populations $p_n$ to the thermal distribution $p_n^{\rm th}$ to confirm that, after being in contact with the engineered bath for a sufficiently long time, the motional mode reaches a thermal steady state (see End Matter). We note that at elevated temperatures ($n_{\rm ss}\gtrsim 4$), phonon-population measurements using the blue-sideband drive are more susceptible to systematic fluctuations in the probe dynamics, such as those caused by laser power and trap-frequency instabilities. These effects can lead to inaccurate estimations of $p_n$, particularly for large $n$, and thus of $n_{\rm ss}$.

\subsection*{Controlled heating of non-COM motional mode} \label{app_noncom}

It is well known in the ion-trapping community that radiative electric-field signal appears approximately uniform across an ion chain and, in the ideal limit, couples only to the center-of-mass (COM) mode. Due to orthogonality of the motional modes, exciting non-COM modes (including the out-of-phase mode used for vibrational encoding in the open-system simulation experiments in the main text \cite{so2025fT}) requires spatial gradients of the electric field \cite{wineland1997issue,brownnutt2015noise,james2000noise}. For this reason, it requires a much higher amplitude of electric-field signal (in this work, approximately 20 times greater in peak voltage) to induce the same heating on the non-COM modes as on the COM mode in the experiments demonstrating thermal-reservoir control in the main text \cite{so2025fT}, using our broadcasting method.

We note that the motional modes of ion crystals usually span frequencies from hundreds of kHz to a few MHz, corresponding to wavelengths on the order of meters, much longer than a typical ion chain. Therefore, the broadcast electric signal operates in the near-field regime, where the antenna geometry significantly influences the field profile experienced by the ion chain. In this work, we utilize a simple helical coil antenna (diameter 45 mm, length 56 mm, and pitch 8 mm), placed approximately 210 mm from the ions, to broadcast the electric-field signals and empirically measure the resulting heating rates on the ions. However, since exciting higher-order, non-COM modes requires spatial overlap between the applied electric-field profiles and the corresponding mode eigenvectors, one can optimize the antenna design and placement by using electric-field simulation tools, such as finite-element analysis, to improve the efficiency of heating these modes. We note that the trap electrodes themselves can also be used as electric-field broadband antennas: because of their close proximity to the ions, they can efficiently couple to higher-order modes and even mediate coupling between two modes, as recently investigated in Refs.~\cite{hou2024swapcool,hou2024modulate}. Moreover, a better stabilization of the laser power and trap frequency would further improve the precision with which the temperature and dissipation rate can be controlled.

\begin{figure}[!t]
\includegraphics[width=0.48\textwidth]{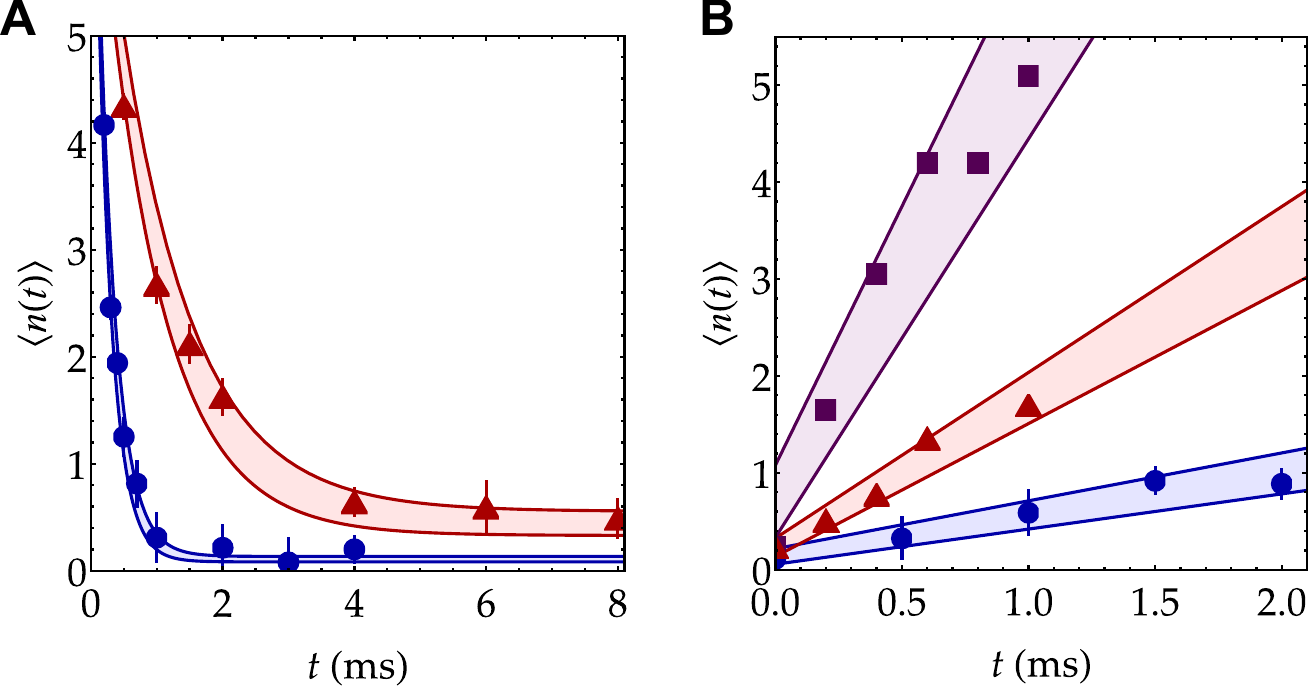}
\vspace{-2 em}
\caption{{\bf Tunability of cooling and heating rates.} (A) Phonon-number dynamics under cooling-only conditions. Red triangles and shaded band are associated with $\gamma_c = 1.00\pm0.10$/ms, while blue circles and band correspond to $\gamma_c=4.03\pm 0.31$/ms. In both cases, the intrinsic heating rate of the targeted mode is $\gamma_h=0.43\pm 0.07$/ms. The curves are fitted exponential functions with time constants $1/\gamma_c$ and the steady states $\braket{n(t\rightarrow\infty)}=\gamma_h/\gamma_c$. (B) Phonon-number dynamics under heating-only conditions. Blue circles and shaded band correspond to the fitted heating rate of the radial center-of-mass mode, $\gamma_h=0.43\pm 0.07$/ms, measured without any engineered heating. Red triangles and shaded band correspond to $\gamma_h=1.54\pm0.17$/ms, while purple squares and shaded band correspond to $\gamma_h=4.73\pm0.62$/ms, both achieved using broadcast electric-field signals. These measured values form part of the calibration procedure for Fig.~2 in the main text \cite{so2025fT}.}
\label{Fig_tunable}
\vspace{-1em}
\end{figure}

\subsection*{Independent cooling and heating rates} \label{app_tune}
To demonstrate the independent tunability of the cooling and heating rates, we include representative evolutions of $\braket{n(t)}$ under cooling-only and heating-only conditions, corresponding to the Fig.~2 data of the main text \cite{so2025fT}, in Fig.~\ref{Fig_tunable}A-B, respectively. The quantity $\braket{n(t)}$ is calculated from the phonon-number distribution $p_n$ obtained from the free-parameter fits. To calibrate the reservoir properties, we first set the cooling-beam parameters and characterize the cooling rate $\gamma_c$ by fitting the phonon-number evolution under the cooling-only condition to an exponential function, where the steady-state phonon number is determined by the known intrinsic heating rate of the targeted mode. We then apply engineered heating to the ion chain in the absence of cooling and extract the heating rate from a linear fit to the phonon-number evolution, where the fitted rate $\gamma_h$ also includes the intrinsic heating rate of the mode. The cooling and heating rates together determine the temperature of the engineered reservoir when both processes are applied simultaneously, as shown in Fig.~2 of the main text \cite{so2025fT}. Although these processes are implemented simultaneously, they are generated independently in our setup via distinct physical processes, so we do not expect significant cross-effects between them, even when operating at higher amplitudes. However, for this protocol to be valid, we must ensure that the spatial extent of the ion's wavefunction, associated with $\braket{n(t)}$, stays within the Lamb-Dicke regime for resolved-sideband cooling to remain effective. Otherwise, Doppler cooling may be used instead.

\subsection*{All-laser thermal-bath engineering} \label{app_alllaser}
In this section, we introduce an alternative thermal-reservoir engineering method using resolved-sideband drives with stochastic phases. Although this method does not offer any experimental advantage over the setup used in the main text \cite{so2025fT}, it is still interesting from a theoretical perspective. 

To demonstrate the physical mechanism of this method, consider a system consisting of a coolant ion and a motional mode. The coolant ion is modeled as a spin-1/2 system, defined by two energy levels $\ket{g}$ and $\ket{e}$, where $\ket{e}$ is a metastable state with lifetime $1/\Gamma$. We note that, in practical qubit-based systems, $\ket{g}$ and $\ket{e}$ may represent the qubit encoded in $\ket{\!\uparrow}_z$ and $\ket{\!\downarrow}_z$ themselves, and the lifetime of $\ket{e}$ can be induced by a qubit-reset operation via optical pumping into $\ket{g}$. For consistency, the spin lowering and raising operators, $\sigma^-$ and $\sigma^+$, are defined in the $\{\ket{g},\ket{e}\}$ basis in the following. A pair of red- and blue-sideband drives is applied to the coolant ion at effective Rabi frequencies $\Omega_r$ and $\Omega_b$, respectively.

When each sideband drive is applied independently to the coolant ion, the red-sideband drive cools the motional mode to $\bar{n}=0$, whereas the blue-sideband drive heats it toward $\bar{n}\to\infty$. A seemingly natural conclusion is that the desired dissipator in Eq.~(1) 
of the main text \cite{so2025fT} could be realized by simply applying both sideband drives simultaneously. However, a closer analysis shows this is not the case. The Hamiltonian describing the coherent effects of the two sideband drives can be written as a generalized red-sideband Hamiltonian:
\begin{eqnarray}
    H_{\rm GRS}= K\sigma^++{\rm h.c.},\;\;K&=&\Omega_ra+\Omega_ba^\dagger.     
\end{eqnarray}
In Ref.~\cite{kienzler2015thesis}, it has been shown that, when $\Gamma\gg\Omega_r,\Omega_b$, the dynamics of the system can be effectively captured by a Lindblad master equation with a single jump operator $c=K$, which yields a steady state $\ket{\psi_s}$ satisfying $K\ket{\psi_s}=0$. It follows that $\bra{\psi_s}[K^\dagger,K]\ket{\psi_s}=|K^\dagger \ket{\psi_s}|^2=\Omega_r^2-\Omega_b^2$. In the case of $\Omega_r\leq\Omega_b$, $\ket{\psi_s}$ is non-physical, and the system cannot equilibrate because there is no valid solution for $\ket{\psi_s}$. When $\Omega_r>\Omega_b $, $\ket{\psi_s}$ is a squeezed vacuum state, characterized by $r=\text{arctanh}(\Omega_b/\Omega_r)$ and $\theta=0$.

\begin{figure}
\includegraphics[width=0.4\textwidth]{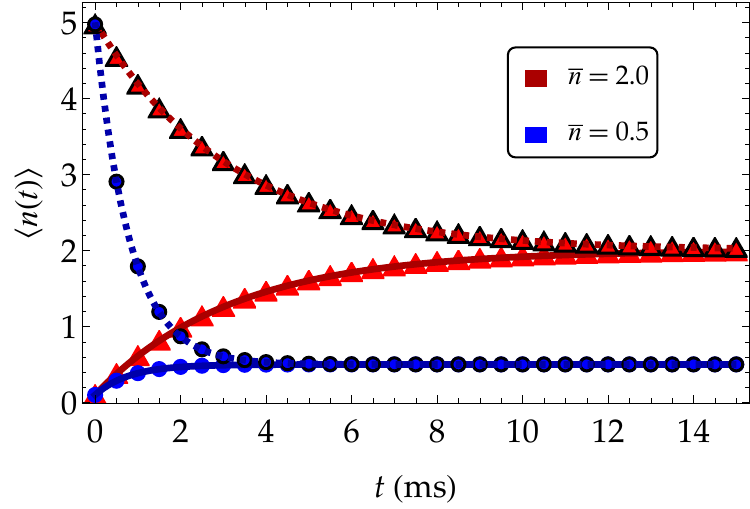}
\vspace{-1 em}
\caption{\textbf{Phonon-number dynamics from the engineered thermal bath based on the all-laser scheme.} Dynamics of $\braket{n(t)}$ obtained by numerically solving Eq.~\eqref{eq_lind_stoch} (triangular and circular dots) and using analytical formula in Eq.~\eqref{eq_ndynamics_2} (solid and dashed curves). Four different setups are implemented with the combinations of $n_0=\{0.1,\;5\}$ and $n_{\rm ss}=\{0.5,\;2\}$. Relevant parameters (in $2\pi\,\times$ kHz) are $\Omega_b=5$ and $\Gamma=10^3$ with the bosonic-space cutoff of $N_c=40$ and discrete time step of $\tau=0.1$ ms (the specific value of $\Omega_r$ is determined from $n_{\rm ss}$ and other parameters for each trajectory).}
\label{Fig_alllaser}
\vspace{-1em}
\end{figure}

To achieve a thermal steady state with controllable temperature, the two sideband drives must therefore be applied incoherently. One possible realization is to apply a stochastic blue-sideband drive, described by:
\begin{eqnarray}
    H_{\rm BS}&=&  \Omega_b(\sigma^+a^\dagger e^{i\phi(t)}+\rm h.c.)\nonumber\\
    &=& \Omega_b(A_1\cos\phi(t)+A_2\sin\phi (t)),
    \label{eq_Hbs}
\end{eqnarray}
where  the values of $\{\phi(t)\}$ are obtained as independent and identically distributed samples from  ${\rm Unif}[0,2\pi)$, $A_1= (a^\dagger \sigma^+ + {\rm h.c.}),$ and $A_2=-i(a^\dagger \sigma^+-\rm h.c.)$. Experimentally, it can be implemented by assigning random constant phases drawn from ${\rm Unif}[0,2\pi)$ in discrete intervals of duration $\tau$. Taking into account the fact that $\ket{e}$ is metastable, the dynamics of the system can be described by a Lindblad master equation:
\begin{eqnarray}
        \partial_t\rho&=&-i[H_{\rm BS},\rho]+ \mathcal{L}_0[\rho],\nonumber\\
        \mathcal{L}_0[\rho]&=&-i[H_{\rm RS},\rho]+\Gamma\mathcal{D}_{\sigma-}[\rho],
        \label{eq_lind_stoch}
\end{eqnarray}
where $H_{\rm RS}=\Omega_{r}(\sigma^+a+{\rm h.c.})$
is the usual red-sideband Hamiltonian.
Using the result of Ref.~\cite{chenu2017noise}, for $\Gamma\gg\Omega_b$, Eq.~\eqref{eq_lind_stoch} can be approximated to the second order as:
\begin{equation}
    \partial_t \rho = \mathcal{L}_0[\rho] - \Omega_b^2\int_0^tC(t,t') dt' \sum_{i=1}^2{\left[A_i\left[A_i^\dagger(t,t'),\rho\right]\right]},
    \label{eq_lind_order2}
\end{equation}
with $A_i(t,t')=\exp(\mathcal{L}_0(t'-t))A_i$. We note here that $A_i$ evolves like a density matrix. With $\Gamma\gg\Omega_r$, the effect of $\mathcal{L}_0$ is dominated by the dissipation term. Using the explicit form of $A_i$ defined in Eq.~\eqref{eq_Hbs}, we approximately get: 
\begin{equation}
    A_i(t,t')\sim \exp(-\Gamma(t-t')/2)A_i.
    \label{eq_form_Ai}
\end{equation}
The two-time correlation function  $C(t,t')$ can be computed in terms of $\tau$:
\begin{equation}
    C(t,t')=\frac{1}{2}\max(0,1-|t-t'|/\tau).
    \label{eq_2time_C}
\end{equation}
Assuming $\tau$ is much smaller than the dynamical timescale ($\tau\ll t$), the integration limit of Eq.~\eqref{eq_lind_order2} can be extended to $t\rightarrow\infty$. Inserting Eqs.~\eqref{eq_form_Ai} and \eqref{eq_2time_C} into Eq.~\eqref{eq_lind_order2}, the evaluation of the integral gives:
\begin{eqnarray}
     \partial_t\rho&=& \mathcal{L}_0[\rho]+\gamma_b\left(\mathcal{D}_{a\sigma^-}\left[\rho\right]+\mathcal{D}_{a^\dagger\sigma^+}\left[\rho\right]\right),\nonumber\\
     \gamma_b&=& \frac{8\Omega_b^2}{\Gamma^2\tau}\left(\frac{\Gamma\tau}{2}-1+e^{-\Gamma\tau/2}\right).
        \label{eq_lind_hc_full}
\end{eqnarray}

Under the limit of $\Gamma\gg \Omega_r,\gamma_b$, the spin part can be effectively eliminated to obtain the following Lindblad equation \cite{kienzler2015thesis}:
\begin{equation}
\partial_t\rho=\gamma_r\mathcal{D}_a\left[\rho\right]+\gamma_b\mathcal{D}_{a^\dagger}\left[\rho\right], \:\: \gamma_r=4\Omega_r^2/\Gamma.
\label{eq_lind_hc_eff}
\end{equation}
The corresponding Heisenberg equation for $\braket{n}$ yields the solution:
\begin{equation}
    \braket{n(t)}=n_{\rm ss}+(\braket{n(t=0)}-n_{\rm ss})e^{-\gamma't},
    \label{eq_ndynamics_2}
\end{equation}
where $\gamma'=\gamma_r-\gamma_b$ and $n_{\rm ss}=\gamma_b/\gamma'$ are the effective equilibration rate and steady-state average phonon population that defines the bath temperature, respectively. 

The equivalence between Eqs.~\eqref{eq_lind_hc_full} and \eqref{eq_lind_hc_eff} can be intuitively understood in the following way. Since $\Gamma$ is large, the coolant ion is kept at $\ket{g}$. Therefore, the red-sideband drive effectively induces a bosonic jump operator $a$. Since the quantum jumps given by the collapse operator $a\sigma^-$ are unlikely to occur because $\ket{g}$ vanishes under the action of $\sigma^-$, the term associated with $a^\dagger\sigma^+$ will be the dominant effect, effectively leading to bosonic heating. Here, the cooling rate can be controlled by the red-sideband Rabi frequency $\Omega_r$, while the heating rate is controlled by that of the stochastic blue-sideband drive $\Omega_b$. In practice, a target temperature $n_{\rm ss}$ can be realized by first fixing $\Omega_b$ and then computing the required $\Omega_r$ from the expressions for $\gamma_b$ and $\gamma_r$ in Eqs.~\eqref{eq_lind_hc_full} and  \eqref{eq_lind_hc_eff}. 

To test the effectiveness of the scheme, we numerically integrate the full Lindblad equation in Eq.~\eqref{eq_lind_stoch} with experimentally implementable parameters (see the caption of Fig.~\ref{Fig_alllaser}). We generate four sets of average phonon-number evolutions with initial temperature $n_0=\{0.1,\;5\}$ and target temperature $n_{\rm ss}=\{0.5,\;2\}$. The stochastic phases are implemented by numerically generating random numbers in each discrete time interval $\tau$ for each trajectory, and we average over $N=50$ stochastic trajectories for each data point. The resulting dynamics of the average boson number $\braket{n(t)}$ are plotted in Fig.~\ref{Fig_alllaser}. The numerical results are consistent with the analytical formula in Eq.~\eqref{eq_ndynamics_2}. The steady states in all four cases are of $\sim 99.99\%$ fidelity with respect to the thermal states with the corresponding target temperatures.

\begin{figure}[!t]
\includegraphics[width=0.49\textwidth]{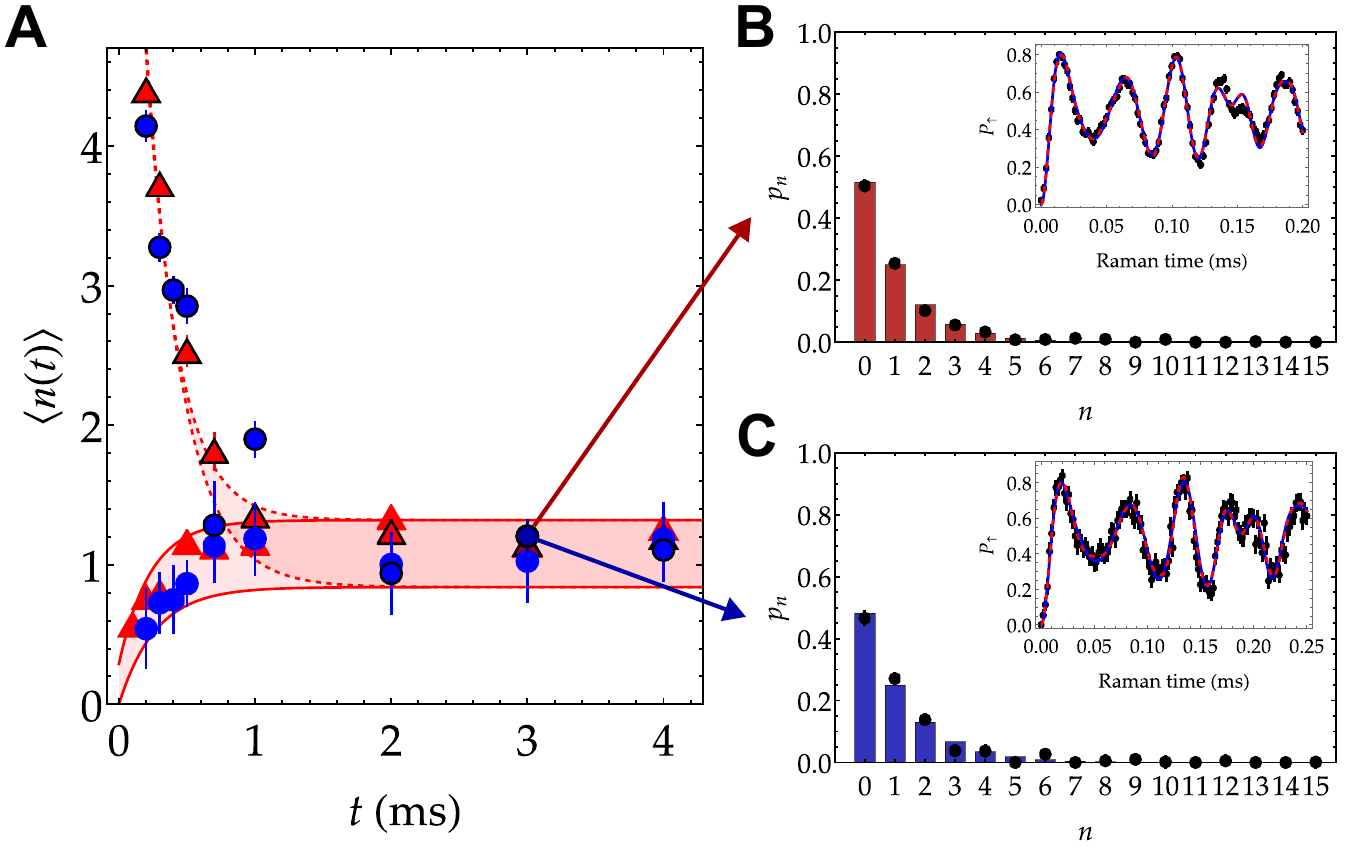}
\vspace{-2 em}
\caption{{\bf Thermal-bath engineering with and without temporal discretization in randomized-phase displacement.} (A) Phonon-number dynamics from interacting with the engineered thermal bath. Red triangles and curves correspond to randomized-phase displacement heating with temporal discretization, whereas blue circles and curves correspond to the case without temporal discretization. The curves are exponential functions with maximum and minimum time constants $1/\gamma_c$ and steady states given by $\braket{n(t\rightarrow\infty)}=\gamma_h/\gamma_c$ with $\gamma_c=4.03\pm 0.31$/ms and $\gamma_h=4.28\pm 0.62$/ms. Data points with black outlines and dashed curves correspond to when the system starts from the Doppler-cooled temperature ($\braket{n(t=0)} \approx 7$), while data points with no outline and solid curves describe the realizations when the system is initialized from $\braket{n(t=0)}\approx0.1$. Darkened red and blue data points at $t=$ 3 ms in (A) are also used in (B) and (C), respectively, for the steady-state phonon-number distributions. Each inset in (B) and (C) corresponds to the spin dynamics of the steady-state system undergoing the probe blue-sideband drive for the phonon-number distribution measurement (black data points in the main figure). Bar charts represent the best-estimated thermal state bounded by the estimated steady states: (B) $0.94\,\pm\,0.05$, and (C) $1.08\,\pm\,0.08$. Dashed red and solid blue curves in the insets describe the dynamics of the fitted phonon-number distribution and the estimated thermal state, respectively.}
\label{Fig_dynamics_undiscretize}
\vspace{-1em}
\end{figure}

\subsection*{Finite-temperature bath engineering using randomized-phase displacement without temporal discretization}

In the main text \cite{so2025fT}, we demonstrate that finite-temperature baths can be engineered by applying laser cooling concurrently with motional heating generated by randomized-phase displacement drives, implemented via temporal discretization and averaging over many realizations. Here, we show that, in the presence of intrinsic motional dephasing (in our system, $\gamma_m $ = 2$\pi\,\times$ 80 Hz) and engineered cooling, applying a displacement drive to a motional mode with a shot-to-shot randomized direction can also yield thermal steady states and dynamics consistent, within our experimental resolution, with low-temperature Lindblad thermal baths ($n_{\rm ss}\lesssim1$). As shown in Fig.~\ref{Fig_dynamics_undiscretize}, the blue data follow the same dynamics as in Fig.~2A of the main text \cite{so2025fT} for $\gamma_h=4.28\,\pm\,0.62$/ms, starting from both a very low temperature ($\braket{n(t=0)}\approx 0.1$) and a Doppler-cooled temperature ($\braket{n(t=0)}\approx 7$). This suggests that, for a single weak displacement drive, motional dephasing suppresses coherent buildup on the relevant thermalization timescale set by the cooling process, while shot-to-shot randomization effectively samples the motional phase space, yielding an approximately thermal state. Indeed, the steady states given by the stochastic drives with (Fig.~\ref{Fig_dynamics_undiscretize}B) and without temporal discretization (Fig.~\ref{Fig_dynamics_undiscretize}C) agree well within our experimental resolution.

\subsection*{Experimental protocol for finite-temperature LVC models}

The following summarizes the experimental protocol for single-mode LVC dynamics (see Refs.~\cite{so2024electrontransfer,so2025multimode} for procedures other than the initial displaced thermal-state preparation and controlled motional heating). The experiment begins by cooling the ion chain near its motional ground state (via Doppler cooling on both species followed by resolved-sideband cooling on the qubit ion), and then optical pumping the qubit to $\ket{\downarrow}_z$. A subsequent $\pi/2$ pulse on the qubit ion maps the $z$ spin basis of the desired LVC Hamiltonian onto the $y$ basis. In this basis, we displace the motional state by $-g/2\omega$ with a spin-dependent optical-dipole force to prepare a near-zero-temperature donor vibronic state ($\approx \ket{D}\otimes\mathcal{D}(-g/2\omega)\ket{0}$, where $\mathcal{D}(\alpha)$ is a displacement operator and $\ket{n}$ represent Fock states).
 
Chemical processes, such as charge transfer, typically occur in thermal equilibrium with their surroundings. It is therefore essential to initialize the donor vibronic state in thermal equilibrium with the bath. To achieve this, prior to the excitation-transfer simulation, we simultaneously apply (\emph{i}) the 355 nm laser tones to generate $H_{\rm unc}\equiv H - V\sigma_x$ on \Yb ion, (\emph{ii}) the 435 nm sympathetic cooling on \Ybc ion, and (\emph{iii}) the motional heating signal for $t_{\rm prep} \gtrsim 4 \times 1/\gamma$, where $\gamma$ is the dissipation rate used in this study. This procedure is equivalent to evolving the system under Eq.~(1) in the main text \cite{so2025fT} with $H \rightarrow H_{\rm unc}$. Without electronic coupling, the spin state initially prepared in the donor state does not evolve, while the motional state evolves to equilibrate with the bath, reaching a thermal displaced steady state $\ket{D}\bra{D}\otimes \rho_{-}$, where $\rho_{-}=\sum_n e^{- n\omega/k_B T}\ket{n_-}\bra{n_-}$ is a thermal state with temperature $k_B T\approx \omega/\log(1 + 1/\bar{n})$, and $\ket{n_{\pm}}=\mathcal{D}(\pm g/2\omega)\ket{n}$ are displaced Fock states (see the Supplementary Materials for Ref.~\cite{so2024electrontransfer}).

\begin{figure}[t!]
\includegraphics[width=0.48\textwidth]{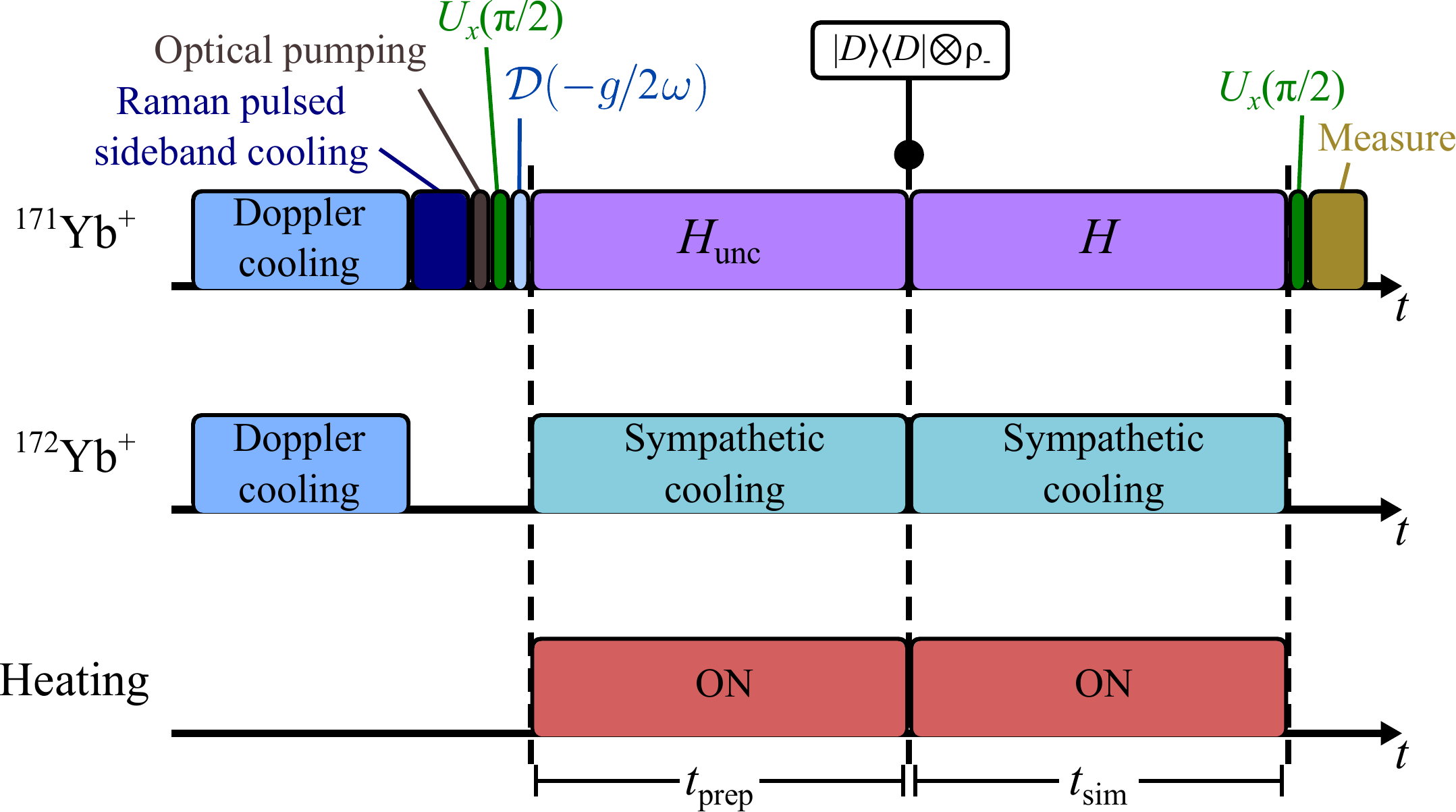}
\caption{{\bf Experimental protocol for finite-temperature single-mode LVC dynamics with a dual-species ion chain.} The preparation steps set the vibrational mode to the desired initial displaced thermal state, $\ket{D}\bra{D}\otimes\rho_-$. The experimental sequence is the same for two-mode simulations, except for an additional displacement step for the second mode.}
\label{Fig_expsequence}
\vspace{-1em}
\end{figure}
 
To simulate excitation transfer from the donor site to the acceptor site, we then turn on the 355 nm laser tone that generates the electronic-coupling term on the qubit ion, creating the total Hamiltonian $H = H_{\rm unc} + V\sigma_x$. After evolving for time $t$, a final $\pi/2$ pulse rotates the state from the $y$ basis back to the $z$ qubit basis, and we measure the donor-state survival probability via state-dependent fluorescence. Calibrations for $H_{\rm unc}$ and $H$ follow Ref.~\cite{so2024electrontransfer}, while controls of the dissipative terms follow the main text \cite{so2025fT}. 
For two-mode simulations, we apply the displacement operation on the motional state of both modes before the simultaneous application of the $H_{\rm unc}$ laser tones, cooling beams, and motional-heating signal.

\subsection*{Transfer-rate definition}

In the main text, we quantify both the equilibration timescale and the net transferred population by the inverse lifetime of the donor-site excitation during the excitation-energy transfer process, $k_T=\frac{\int_{0}^{t_\text{sim}} P_D(t)dt}{\int_{0}^{t_\text{sim}} t P_D(t)dt} - \frac{2}{t_\text{sim}},$
where $t_{\rm sim}$ is the simulation duration, and $P_D(t)=(\braket{\sigma_z(t)}+1)/2$. This quantity is often referred to as the transfer rate, but it may be more intuitive to think of it as the transfer efficiency, as it depends both on the timescale and the steady state of the transfer process. For exponentially decaying dynamics with no residual population in the donor site, this transfer-rate definition reduces, at sufficiently long times, to the inverse of the decay time constant, consistent with the first-order rate law used in chemical kinetics \cite{schlawin2021electrontransfer,padilla2025delocalizedexcitationtransferopen}. Accordingly, this metric is most appropriate for dynamics that approach a steady state. We note that the term $-\frac{2}{t_{\rm sim}}$ is included to remove the background contribution associated with the finite-time evaluation of the non-zero steady-state donor population in the transfer-rate calculations, without altering the characteristic features of the resulting transfer-rate spectra \cite{so2024electrontransfer,so2025multimode}.

\subsection*{Numerical calculations for finite-temperature excitation transfer}\label{app_numeric}

We employ a Python package built on QuTiP \cite{johansson2013qutip} to numerically simulate excitation-transfer dynamics of open-system LVC models for comparison with experimental data \cite{so2024electrontransfer,so2025multimode}. To account for experimental imperfections, we extend the simulation of Eq.~(1) in the main text \cite{so2025fT} to include additional decoherence channels:
\begin{eqnarray}
\frac{\partial\rho}{\partial t}\!\!\!&=&\!\!\!-i[H,\rho] +\! \sum_{i=1}^2\left\{\gamma_i (\bar{n}_i+1)\mathcal{L}_{a_i}[\rho] + \gamma_i \bar{n}_i \mathcal{L}_{a_i^\dagger}[\rho]\right\} \nonumber \\
&\quad& \quad\quad+\;\gamma_z\mathcal{L}_{\sigma_y}[\rho]+\sum_{i=1}^2\gamma_{m}\mathcal{L}_{a_i^\dagger a_i}[\rho].
\label{eq_master_mod}
\end{eqnarray}
Here, the jump operator $\sigma_y$ with rate $\gamma_z$ models spin dephasing due to laser-power fluctuations in the rotated spin basis ($z \leftrightarrow y$), while the jump operators $a_i^\dagger a_i$ with a common rate $\gamma_{m}$ describe motional dephasing caused by trap-frequency fluctuations \cite{so2025multimode}. By comparing the numerical results with the experimental data, we extract $\gamma_z/2\pi = 7$ Hz and $\gamma_{m}/2\pi = 80$ Hz, both much smaller than the engineered terms. For single-mode calculations, $\gamma_i\equiv\gamma$ and $\bar{n}_i\equiv\bar{n}$.

\subsection*{High-temperature charge transfer} \label{app_opt}

\begin{figure}[t!]
\includegraphics[width=0.48\textwidth]{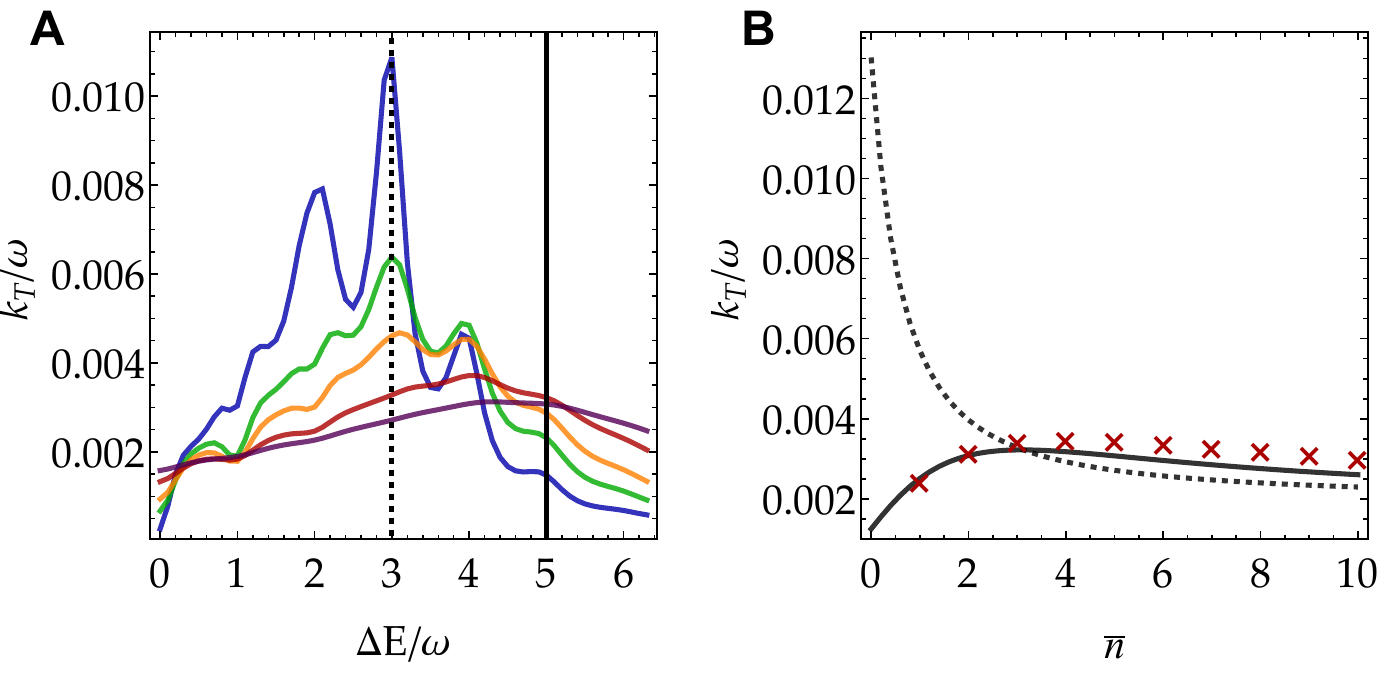}
\vspace{-2 em}
\caption{{\bf Temperature effects on charge transfer.} (A) Transfer rate versus donor-acceptor energy gap. Blue, green, orange, red, and purple solid lines show numerical results of the transfer rate with $(V,\;g)=(0.2,\;1.1)\omega$ in contact with a thermal reservoir characterized by $\bar{n}=0.15$, 0.80, 1.50, 3.00, and 5.00, respectively, at a dissipation rate of $\gamma = 0.036\omega$. These solid lines also account for experimental imperfections with $(\gamma_z,\;\gamma_m)  = (0.001,\;0.016)\omega$, enabling a direct comparison to Fig.~4 of the main text. (B) Transfer rate versus average phonon number of the thermal bath. Dashed and solid black lines correspond to $\Delta E = 3\omega$ and $\Delta E = 5\omega$, respectively (vertical lines in panel (A)). Red crosses are FGR rates with an empirical prefactor of $A$ = 1.18.}
\label{Fig_Temp}
\vspace{-1em}
\end{figure}

In the main text \cite{so2025fT}, we present strongly adiabatic charge-transfer rate spectra for two average phonon numbers of the thermal bath, $\bar{n}=0.15$ and $\bar{n}=0.80$, which correspond to $k_B T\approx0.49\omega$ and $k_B T\approx1.23\omega$, respectively. At these temperatures ($k_B T\sim\omega$), the vibrational mode still behaves quantum mechanically  \cite{schlawin2021electrontransfer}. However, as Fig.~\ref{Fig_Temp}A-B illustrates, raising the environment temperature progressively flattens the transfer rate spectrum, rendering it featureless with respect to the donor-acceptor energy gap (significantly decreased transfer rates at low $\Delta E$ and slightly enhanced transfer rates at high $\Delta E$). This suggests the transition of charge-transfer behaviors into the classical regime ($k_B T\gg\omega$), where the vibrational mode is no longer quantized \cite{schlawin2021electrontransfer}. 

While the temperature-induced flattening may seem obvious, a closer look at systems with large $\Delta E$ provides insight into why a slight rate enhancement occurs. In this $\Delta E$ regime, the upper and lower adiabatic surfaces closely resemble the uncoupled donor and acceptor surfaces (see End Matter). Consequently, the transfer rates can be approximated by Fermi’s golden-rule (FGR) transition rates $k_{\rm FGR}=A|V|^2 {\rm FC}(g)$, where ${\rm FC}(g)\equiv\sum_{m}p_m{\rm FC}_{m,m+n}(g)$ is the total population-weighted Franck-Condon overlap between donor and acceptor vibronic levels, which depends on displacement ($g$) and temperature with $n=\Delta E/\omega$ being an integer. Since FGR analysis here does not explicitly account for dissipation and non-zero donor-acceptor hybridization, we incorporate these effects through an empirically determined prefactor $A$, which makes the FGR rates agree qualitatively with the numerical calculations (see Fig.~\ref{Fig_Temp}B). 

The similarity between the numerical results and the modified FGR predictions indicates that the slight enhancement of transfer rates at large $\Delta E$ also arises from the temperature-dependent distribution of population, and that an optimal temperature exists for a given $\Delta E$ (set by the Franck-Condon factors). We also note that the deviation between the numerically calculated rates and the modified FGR rates at high temperatures in Fig.~\ref{Fig_Temp}B may be due to incomplete population transfer, where thermal backflow can leave residual population trapped on the donor site at equilibrium \cite{padilla2025delocalizedexcitationtransferopen}, a feature not be captured by the FGR treatment.

\begin{figure}[t!]
\vspace{1 em}
\includegraphics[width=0.4\textwidth]{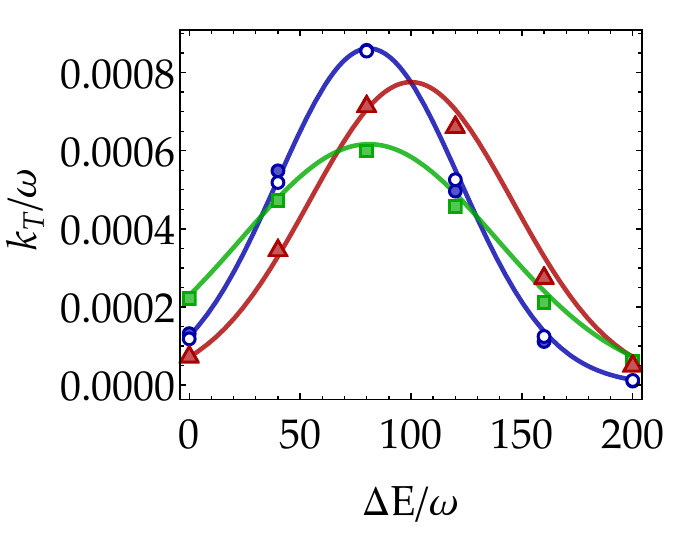}
\vspace{-1 em}
\caption{{\bf Classical nonadiabatic charge transfer.} (A) Transfer rate versus donor-acceptor energy gap. Points and solid lines correspond to numerical calculations and $\sqrt{\pi}k_{\rm M}$ (in units of $\omega$), respectively.  Blue, green, and red data are associated with system-bath parameters $(V,\;g,\;k_B T)=\{(0.2,\;9,\;10.5)\omega,\; (0.2,\;9,\;20.5)\omega,\;(0.2,\;10,\;10.5)\omega\}$, respectively. Open and filled data points correspond to $\gamma=0.2\omega$ and $\gamma=0.4\omega$, respectively. In the numerical calculations, we consider the phonon cutoff $N_c = 230$, and the total evolution time $\omega t_{\rm sim}/(2\pi)=250$.}
\label{Fig_Marcus}
\vspace{-1em}
\end{figure}

However, we note that the transfer rates in the intermediate regime ($\lambda\equiv g^2/\omega\sim\omega$) presented above lie outside the applicability of Marcus theory, which assumes charge-transfer-active vibrational modes with $\lambda\gg\omega$ and the classical limit with $k_B T\gg\omega$ \cite{schlawin2021electrontransfer}.
For completeness, we numerically verify the transfer-rate spectra in the classical nonadiabatic regime, where $\lambda\gg\omega$, $k_B T\gg\omega$, and $\pi^{3/2}|V|^2/(\omega\sqrt{\lambda k_B T})\ll1$, against Marcus theory \cite{marcus1993electron,schlawin2021electrontransfer}, whose rate is given by:
\begin{equation}
   k_{\rm M}=|V|^2\sqrt{\frac{\pi}{\lambda k_B T}}\exp{\left[-\frac{(\lambda-\Delta E)^2}{4\lambda k_B T}\right]}.
\end{equation}
As shown in Fig.~\ref{Fig_Marcus}, the transfer rates from the numerical calculations (point data) in all four cases qualitatively agree with the corresponding Marcus predictions (solid curves) up to an empirically determined prefactor of $\sqrt{\pi}$. This constant prefactor likely arises from differences in transfer rate conventions. We also note that, in this regime, the transfer rate is weakly dependent on the dissipation rate $\gamma$ for $\gamma\lesssim\omega$ (compare open and filled blue circles), consistent with Marcus theory, whose rate has no explicit dependence on $\gamma$.

\vspace{-1 em}
\subsection*{Finite-temperature vibrationally assisted exciton transfer}
\vspace{-0 em}
\begin{figure}[t!]
\vspace{1em}
\includegraphics[width=0.35\textwidth]{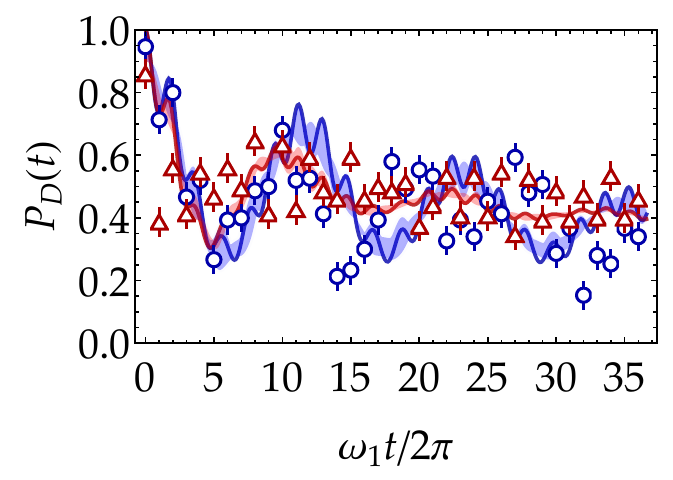}
\vspace{-1em}
\caption{{\bf Dissipative,  finite-temperature single-phonon exchange.} Donor population dynamics for $\Delta E = 0.5\omega_1$, near single-$\omega_2$-phonon exchange resonance. Blue curves and circles show the theoretical and experimental results, respectively, for the system interacting with the $\bar{n}_2=0.02$ bath, while red curves and triangles correspond to the $\bar{n}_2=0.80$ bath. The shaded bands on the theoretical curves correspond to the mean uncertainty ($\pm\,0.20$) in the most sensitive parameters in our setup, the bath temperatures, and imperfect initial-state preparation.}
\label{Fig_singlephonon}
\vspace{-1em}
\end{figure}

To build intuition for the role of temperature in vibrationally assisted exciton transfer, we consider the perturbative regime, where $g_i$ are much smaller than the other parameters in Eq.~(4) in the main text \cite{so2025fT}. Following Ref.~\cite{so2025multimode}, the eigenstates of the total system without the vibronic couplings are $\ket{e_\pm,n_1,n_2}$. Here, $\ket{e_\pm}$ are mixtures of $\ket{\uparrow}_z$ and $\ket{\downarrow}_z$ with $\ket{e_+}\rightarrow \ket{\uparrow}_z$ for $V/\Delta E \rightarrow 0$ ($\Delta E > 0$), and $n_1$ and $n_2$ describe the Fock states of the two harmonic oscillators.
In this eigenbasis, the vibronic coupling terms can be written as the following perturbation:
\begin{equation}
    H_{p} = \!\sum_{i=1}^2 \frac{g_i}{2\sqrt{\left(\frac{\Delta E}{2}\right)^2\!+\!V^2}}\left(\frac{\Delta E}{2}\tilde{\sigma}_z\!-\!V\tilde{\sigma}_x\right)(a_i+a_i^\dagger).
\end{equation}

\begin{figure}[t!]
\vspace{1 em}
\includegraphics[width=0.4\textwidth]{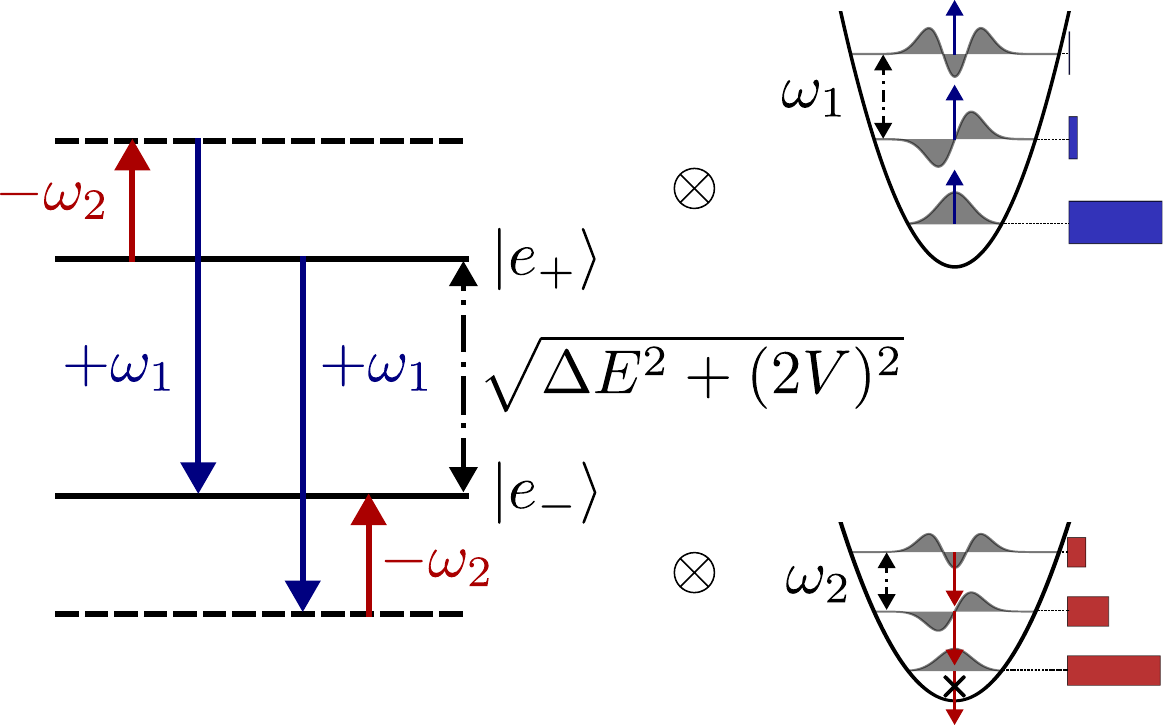}
\caption{{\bf Thermally activated transfer pathways.} Second-order vibrationally assisted exciton transfer, where weak vibronic coupling coherently drives transitions between eigenstates of the electronically coupled system, $\ket{e_+}$ and $\ket{e_-}$, via simultaneous excitation of one $\omega_1$ phonon and de-excitation of one $\omega_2$ phonon. Bars next to the harmonic-oscillator levels indicate the corresponding steady-state populations for ($\bar{n}_1,\;\bar{n}_2$) = (0.10, 0.80). This second-order process is forbidden for the population in $\omega_2$-oscillator's ground state. The equilibration due to dissipation is not shown. We note that the dashed lines denote virtual states (electronically, can be either $\ket{e_+}$ or $\ket{e_-}$ on the same line) with the colored solid arrows indicating changes in phonon numbers; however, they do not represent the true energy levels of the virtual states.}
\label{Fig_thermallyactscheme}
\vspace{-1em}
\end{figure}

For $\Delta E>0$, Fermi's golden rule gives a single-phonon excitation rate for each level pair that is proportional to $p_{n_i}(n_i+1)$, implying that higher temperatures speed up the transfer dynamics \cite{gorman2018VAET,so2025multimode}. Since hotter systems feature more $p_{n_i}\neq0$, they also activate more pairwise-transition channels, which accelerate decoherence during the evolution. We confirm these effects in both numerical calculations and experiment (see Fig.~\ref{Fig_singlephonon}). However, because the transfer-rate definition in the main text accounts for population both leaving and returning to the donor site during the dynamics, higher temperatures cause the excitation to spend a significant amount of time on the donor site, yielding a lower net transfer rate (see Fig.~4 of the main text \cite{so2025fT}).

Although similar intuition applies to other vibrationally assisted exciton-transfer processes, including the higher-order phonon exchanges \cite{so2025multimode}, the mechanism underlying the thermal activation of the mixed-mode resonance, corresponding to excitation of the $\omega_1$ mode and de-excitation of the $\omega_2$ mode, in Fig.~4 of the main text \cite{so2025fT} is different. As shown in Fig.~\ref{Fig_thermallyactscheme}, this mixed-mode process involves interfering pathways with excitation/absorption of one $\omega_1$ phonon and de-excitation/emission of one $\omega_2$ phonon, in either order. Specifically, the total vibronic system evolves through two virtual states in four following coherent pathways:

\noindent$\ket{e_+,n_1,n_2}$$\rightarrow$$\ket{e_+,n_1+1,n_2}$$\rightarrow$$\ket{e_-,n_1+1,n_2 -1},$

\noindent$\ket{e_+,n_1,n_2}$$\rightarrow$$\ket{e_+,n_1,n_2-1}$$\rightarrow$$\ket{e_-,n_1+1,n_2 -1},$

\noindent$\ket{e_+,n_1,n_2}$$\rightarrow$$\ket{e_-,n_1+1,n_2}$$\rightarrow$$\ket{e_-,n_1+1,n_2 -1},$

\noindent$\ket{e_+,n_1,n_2}$$\rightarrow$$\ket{e_-,n_1,n_2-1}$$\rightarrow$$\ket{e_-,n_1+1,n_2 -1}.$

\noindent When both temperatures of the modes are low, their population resides mainly in the ground state, suppressing the de-excitation/emission process. Raising the local temperature of the oscillator at frequency $\omega_2$ populates excited states, enabling de-excitation and thereby activating the transfer pathways based on constructive interference.

\begin{figure}[t!]
\vspace{1em}
\includegraphics[width=0.35\textwidth]{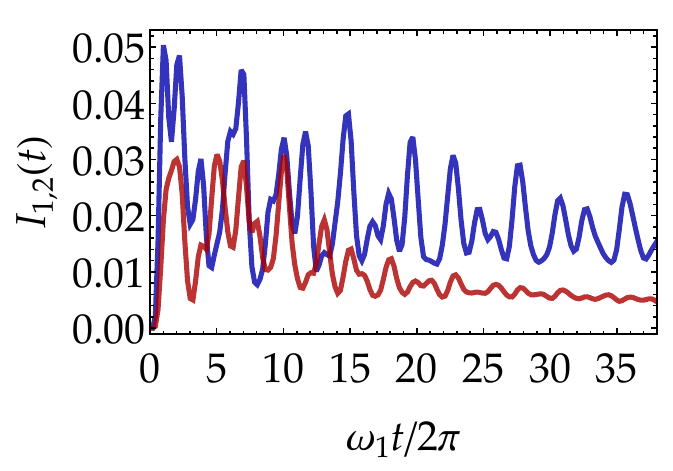}
\vspace{-1em}
\caption{{\bf Mixed-mode coherence during LVC evolution.} Time evolution of the mutual information between the two vibrational modes for $\Delta E = 0.3\omega_1$, near the mixed-mode exchange resonance. Blue curve shows the theoretical result for the system interacting with the $\bar{n}_2=0.02$ bath, while red curve corresponds to the $\bar{n}_2=0.80$ bath.}
\label{Fig_mixedmodecoherence}
\vspace{-1em}
\end{figure}

It is important to note that, before the system reaches equilibrium, coherence develops between the two vibrational modes during the excitation-transfer dynamics, even though each mode is individually thermalized by its engineered bath. This coherence arises from their simultaneous coupling to the donor and acceptor states. We confirm this numerically by examining the evolution of the quantum mutual information between the two bosonic modes, given by $I_{1,2}=S(\rho_1)+S(\rho_2)-S(\rho_{12})$, where $S$ is the von Neumann entropy, $\rho_{12}$ is the reduced two-mode phonon density matrix of the full system, and $\rho_1$ and $\rho_2$ are the reduced density matrices of modes 1 and 2, respectively. The $I_{1,2}(t)$ for $\Delta E = 0.3\omega_2$ plotted in Fig.~\ref{Fig_mixedmodecoherence} shows that steady-state coherence exists for both $\bar{n}_2>0$ and $\bar{n}_2\sim0$. In the former case, we attribute the non-zero mutual information primarily to the energy exchange between the donor-acceptor states and the two vibrational mode (mixed-mode). On the other hand, at low temperature, we attribute the coherence buildup to off-resonant processes, given by simultaneous two-phonon energy exchange between the donor-acceptor system and each vibrational mode. We emphasize that mutual information does not distinguish classical and quantum correlations. In our system, numerically accessible entanglement measures, such as negativity, exhibit only a very weak signal, making the detection of genuine quantum entanglement challenging. Nevertheless, this does not preclude entanglement, since a non-zero negativity provides only a sufficient criterion.

\end{document}